\documentclass[12pt,preprint]{aastex}

\def\logg{\ifmmode {\log\;{\rm g}}\else log g\fi}
\def\feh{\ifmmode {\rm [Fe/H]}\else [Fe/H]\fi}
\def\fehm{\ifmmode {\rm [Fe/H]}_{\rm m}\else [Fe/H]$_{\rm m}$\fi}
\def\feho{\ifmmode {\rm [Fe/H]}_{\rm o}\else [Fe/H]$_{\rm o}$\fi}
\def\fehpk{\ifmmode {\rm [Fe/H]}_{\rm pk}\else [Fe/H]$_{\rm pk}$\fi}
\def\fehpa{\ifmmode {\rm [Fe/H]}_{\rm pa}\else [Fe/H]$_{\rm pa}$\fi}
\def\fehki{\ifmmode {\rm [Fe/H]}_{\rm K1}\else [Fe/H]$_{\rm K1}$\fi}
\def\fehkii{\ifmmode {\rm [Fe/H]}_{\rm K2}\else [Fe/H]$_{\rm K2}$\fi}
\def\fehkiii{\ifmmode {\rm [Fe/H]}_{\rm K3}\else [Fe/H]$_{\rm K3}$\fi}
\def\fehai{\ifmmode {\rm [Fe/H]}_{\rm A1}\else [Fe/H]$_{\rm A1}$\fi}
\def\fehaii{\ifmmode {\rm [Fe/H]}_{\rm A2}\else [Fe/H]$_{\rm A2}$\fi}
\def\fehaiii{\ifmmode {\rm [Fe/H]}_{\rm A3}\else [Fe/H]$_{\rm A3}$\fi}
\def\fehaki{\ifmmode {\rm [Fe/H]}_{\rm AK1}\else [Fe/H]$_{\rm AK1}$\fi}
\def\fehakii{\ifmmode {\rm [Fe/H]}_{\rm AK2}\else [Fe/H]$_{\rm AK2}$\fi}
\def\teff{\ifmmode {\rm{T}_{\rm{eff}}}\else T$_{\rm{eff}}$\fi}
\slugcomment{}
 
\shorttitle{Carbon Abundances in Metal-Deficient Stars}
\shortauthors{Rossi et al.}
 
\begin{document}
 
\title{Estimation of Carbon Abundances in Metal-Poor Stars. I.  Application to the 
``Strong G-band'' stars of Beers, Preston, \& Shectman
}

\author{Silvia Rossi\altaffilmark{1,2}}
\affil{Instituto de Astronomia,  Geof\'{i}sica e Ci\^{e}ncias Atmosf\'{e}ricas, Departamento de Astronomia,  Universidade de S\~{a}o Paulo, \\ 
Rua do Mat\~{a}o  1226, 05508-900 S\~{a}o
Paulo, Brazil}
\email{rossi@astro.iag.usp.br}

\author{Timothy C. Beers\altaffilmark{1,2,3,4}}
\affil{Department of Physics \& Astronomy and JINA: Joint Institute for Nuclear
Astrophysics,\\ Michigan State University,
 E. Lansing, MI  48824}
\email{beers@pa.msu.edu} 

\author{Chris Sneden}
\affil{Department of Astronomy, University of Texas, Austin, TX  78712} 
\email{chris@verdi.as.utexas.edu} 

\author{Tatiana Sevastyanenko\altaffilmark{1}}
\affil{Department of Physics \& Astronomy, Michigan State University,
 E. Lansing, MI  48824}
\email{sevastya@pa.msu.edu}

\author{Jaehyon Rhee\altaffilmark{1,2}}
\affil{Center for Space Astrophysics, Yonsei University, Seoul 120-749, Korea, and 
Space Astrophysics Laboratory, \\ California Institute of Technology, MC 405-47, Pasadena, CA 91125}
\email{rhee@caltech.edu}

\author{Brian Marsteller\altaffilmark{1}}
\affil{Department of Physics \& Astronomy and JINA: Joint Institute for Nuclear
Astrophysics,\\ Michigan State University, E. Lansing, MI  48824}
\email{marsteller@pa.msu.edu}
 
\altaffiltext{1}{Visiting Astronomer, Kitt Peak National Observatory.  KPNO is
operated by AURA, Inc., under contract to the National Science Foundation.}
 
\altaffiltext{2}{Visiting Astronomer, Cerro Tololo Interamerican Observatory
CTIO is operated by AURA, Inc., under contract to the National Science
Foundation.}
 
\altaffiltext{3}{Visiting Astronomer, European Southern Observatory, La Silla,
Chile}

\altaffiltext{4}{Visiting Astronomer, McDonald Observatory, University of Texas}

\begin{abstract}
 
We develop and test a method for the estimation of metallicities ([Fe/H]) and
carbon abundance ratios ([C/Fe]) for carbon-enhanced metal-poor (CEMP) stars, based
on application of artificial neural networks, regressions, and synthesis models
to medium-resolution ($1-2$ \AA\ ) spectra and $J-K$ colors. We calibrate this
method by comparison with metallicities and carbon abundance determinations for
118 stars with available high-resolution analyses reported in the recent
literature. The neural network and regression approaches make use of a
previously defined set of line-strength indices quantifying the strength of the
CaII K line and the CH G-band, in conjuction with $J-K$ colors from the 2MASS
Point Source Catalog. The use of near-IR colors, as opposed to broadband $B-V$
colors, is required because of the potentially large affect of strong molecular
carbon bands on bluer color indices. 

We also explore the practicality of obtaining estimates of carbon abundances for
metal-poor stars from the spectral information alone, i.e., without the
additional information provided by photometry, as many future samples of
CEMP stars may lack such data. We find that, although
photometric information is required for the estimation of [Fe/H], it provides
little improvement in our derived estimates of [C/Fe], hence estimates of
carbon-to-iron ratios based solely on line indices appear sufficiently accurate for
most purposes. Although we find that the spectral synthesis approach yields the
most accurate estimates of [C/Fe], in particular for the stars with the
strongest molecular bands, it is only marginally better than is obtained from
the line index approaches.

Using these methods we are able to reproduce the previously-measured [Fe/H] and
[C/Fe] determinations with an accuracy of $\approx 0.25$ dex for stars in the
metallicity interval $-5.5 \le {\rm [Fe/H]} \le -1.0$, and with $0.2 \le (J-K)_0
\le 0.8$. At higher metallicity the CaII K line begins to saturate, especially for 
the cool stars in our program, hence this approach is not useful in some cases.
As a first application, we estimate the abundances of [Fe/H] and [C/Fe] for the
56 stars identified as possibly carbon-rich, relative to stars of similar metal
abundance, in the sample of ``strong G-band'' stars discussed by Beers, Preston,
\& Shectman. 
 
\end{abstract}

 
\section{Introduction}
 
Several recent papers (e.g., Norris, Ryan, \& Beers 1997a; Rossi, Beers, \&
Sneden 1999) have pointed out that the frequency of carbon-enhanced stars in the
Galaxy appears to increase at the lowest metal abundances. Continuation and
expansion of ongoing medium-resolution spectroscopic follow-up of candidate
low-metallicity stars (the HK survey of Beers and colleagues; Beers, Preston, \&
Shectman 1992, hereafter BPSII; Beers 1999; the Hamburg/ESO Survey, HES, of
Christlieb and collaborators; Christlieb 2003) has indicated that the actual
fraction of stars with metallicities [Fe/H] $ \le -2.0$ and carbon enhancements
in excess of [C/Fe] $ \simeq +1.0$ may be even higher than previously
suspected, perhaps as great as 20-25\%.  At the lowest metallicities, e.g., for [Fe/H]
$< -3.5$, the fraction of stars that are carbon enhanced at a level of [C/Fe]
$\ge +1.0$ rises to 40\% (Beers \& Christlieb 2005).  The only two
stars known with [Fe/H] $< -5.0$ {\it both} exhibit extremely high [C/Fe]
ratios (Christlieb et al. 2002; Frebel et al. 2005).

Recently, Cohen et al. (2005) have argued that the fraction of carbon-enhanced,
metal-poor (hereafter, CEMP) stars has been overestimated by previous studies.
They use a sample of some 50 high-resolution spectra of metal-poor stars
selected from the HES to conclude that the fraction of CEMP stars is 14.4\% $\pm$
4\%.  This fraction does not differ at the two-sigma level from previous claims.
Furthermore, the authors of the present paper feel that there are multiple reasons why
this determination might be more properly be considered a lower limit on the
actual frequency, including the procedures used to select metal-poor candidates
from the HES.  The question is of great significance, and needs to be revisited
using larger samples of metal-poor stars with high-resolution abundance
analyses, as well as with stars selected in alternative ways.

Clearly, it is important to understand the astrophysical phenomena responsible
for the high frequency of CEMP stars, and to assess their impact on the early
chemical evolution of the Galaxy. Lucatello et al. (2005b), for instance, have
argued that the large fraction of CEMP stars at low metallicities provides
evidence for an alteration in the Initial Mass Function (IMF) during these
epochs to include a greater number of intermediate mass stars than are formed
from the present-day IMF. The connection, if any, to the significant amounts of
ionized carbon in the intergalactic medium detected in observations toward
distant quasars (e.g., Ellison et al 2000; Pettini et al. 2003), may also hold
important clues to the production of carbon at the earliest times. For
additional details, see the discussion in Beers \& Christlieb (2005).

High-resolution abundance analyses for a number of CEMP stars (Barbuy et al.
1997; Norris, Ryan, \& Beers 1997b; Bonifacio et al. 1998; Aoki et al. 2000,
2001, 2002a,b,c; Norris et al. 2002; Depagne et al. 2002; Sivarani et al. 2004;
Barbuy et al. 2005) indicates that a variety of carbon-production mechanisms may
need to be invoked to account for the observed range of elemental abundance
patterns in these stars (e.g., mass-transfer from former AGB companions,
self-pollution via mixing of CNO-processed material in individual stars, the
possible existence of zero-abundance ``hypernovae'' which may produce large
amounts of carbon, etc.). Many of the CEMP stars have been shown to be members
of binary systems (Preston \& Sneden 2001; Tsangarides, Ryan, \& Beers 2004;
Lucatello et al. 2005a). The majority, but interestingly not all, of the CEMP stars are
associated with neutron-capture-element enhancement (in particular from the
s-process; see Aoki et al. 2003). At least one member of the growing class of
highly r-process-enhanced, metal-poor stars, CS~22892-052 (see Sneden et al.
1996, 2003) also exhibits large C (and N) overabundances relative to the solar
ratios, although the origin of the carbon enhancement and the r-process
enhancement may well be decoupled from one another. It is surely not a
coincidence that the two most iron-deficient stars yet identified, HE~0107-5240
(Christlieb et al. 2002; Christlieb et al. 2004), and HE~1327-2326 (Frebel et
al. 2005) with [Fe/H] = --5.3 and --5.6, respectively, also exhibit carbon
overabundances that are the highest yet reported amongst extremely metal-poor
stars, on the order of [C/Fe] $\simeq +4.0$. Much clearly remains to be learned
about the nature, origin, and evolution of the many classes of CEMP stars in the
Galaxy.

Recent analysis of stellar spectra from the Sloan Digital Sky Survey (Margon et
al. 2002; Downes et al. 2004) has led to the claim that at least 50\% of the
carbon-enhanced stars in the Galaxy are probably dwarfs. The working hypothesis
is that the majority of these stars are the result of mass-transfer episodes
from a now-extinct more massive companion. Totten \& Irwin (1998) have called
attention to the presence of (apparently) intermediate-mass stars, still in
their AGB phase, that may have been stripped from the Sagittarius dwarf galaxy,
and which provide a potentially strong probe on the clumpiness and axial
symmetry of the Galaxy (Ibata et al. 2001).   
 
In order to quantify and understand the varieties and mechanisms for explaining
the origin of CEMP stars it is necessary to develop procedures by which [Fe/H]
and [C/Fe] measurements may be rapidly obtained for as large a number
of stars as possible. Ideally, this should be based on medium-resolution (1-2
{\AA}) spectroscopy, since this information is already available for many
thousands of metal-poor stars from the HK and HES follow-up campaigns, and will
be available for hundreds of thousands of stars as new-generation surveys expand
(e.g., SDSS: York et al. 2000, RAVE: Steinmetz 2003, and SEGUE: The Sloan
Extension for Galactic Understanding and Exploration, which is part of the
extension to the SDSS, known as SDSS-II). 

In this paper we obtain a calibration of methods for rapid analysis of CEMP
stars, and apply it to the sample of ``strong G-band'' stars from the HK survey
noted by BPSII. In \S 2 we describe the selection of the calibration stars and
the acquisition and analysis of the spectroscopic and photometric data used in
this study. In \S 3 we describe the training and application of an Artificial
Neural Network (ANN) approach, a regression approach, and a spectroscopic
fitting approach.  Application of our calibration to
the ``strong G-band'' stars of BPSII is reported in \S 4. In \S 5 we discuss the
distribution of carbon abundances derived for these stars. A summary of our
results and comments on future applications of these techniques is presented in
\S 6.

\section{Selection of Calibration Stars and Observational Data}
 
In the assembly of our calibration sample we have endeavored to include
metal-deficient stars with as wide a range of physical parameters (T$\rm{eff}$,
log g, [Fe/H]) and known carbon abundances as possible. Fortunately, recent
interest in the nature of CEMP stars has greatly increased the number of
suitable calibration stars. We have combined these objects with modern
analyses of additional metal-poor stars in order to better cover the range of
carbon abundances known to exist amongst stars in the halo of the Galaxy (e.g.,
$-1.0 \le {\rm [C/Fe]} \le +4.0$). Although we would have liked to include
significant numbers of dwarfs in this exercise, there are only a handful of such
stars with well-measured carbon abundances that exceed the solar ratio of
[C/Fe], so out of necessity our calibration sample is dominated by subgiants and
giants. 

We also require medium-resolution spectra for each star in the calibration
sample. An initial effort (reported by Rossi et al. 1999) revealed that there
were ranges of the calibration space that needed to be filled in with new
medium-resolution spectroscopy of stars having available high-resolution carbon
abundance determinations. Over the past few years we have acquired these
data during the course of ongoing HK and HES spectroscopic follow-up campaigns.

\subsection{Photometry}

The broadband $V$ magnitudes and $B-V$ colors for our calibration stars are
taken either from the SIMBAD database, or (for the majority of the HK survey
calibration stars) from BPSII, McWilliam et al. (1995), or Beers et al. (2005,
in preparation). It was recognized early on that $B-V$ colors are less than
ideal for use in the estimation of carbon abundances in metal-poor stars, owing
to the (in some cases) rather significant alteration of the observed flux in
these bands due to the presence of molecular CH, CN, and C$_2$ features. This is
particularly true for the most carbon-rich and/or cooler stars.  One might
attempt to use the $HP2$ index, a band-switched measure of the strength of the
absorption line H-$\delta$ (as defined by Beers et al. 1999),
as an alternative temperature indicator. Although this is useful for warmer
stars, e.g., those with T$_{\rm eff} > 5000$ K, for cooler stars $HP2$ weakens
precipitously, is subject to greater measurement errors, and loses some of its
sensitivity to stellar temperature.  Hence, we are
fortunate that we can now take advantage of the reasonably high-quality
near-infrared $JHK$ photometry that has recently become available from the 2MASS
project (Cutri et al. 2003). The application of $J-K$ colors from 2MASS provides
a indicator of stellar temperature that is less sensitive to molecular carbon
bands, and furthermore, is much less dependent on foreground reddening than is
the $B-V$ color.

Table 1 presents the available photometry for our calibration stars.
Column (1) lists the star name. Columns (2) and (3) provide the $V$ magnitude
and $B-V$ color, respectively. In column (4) and (5) we list the $J$ magnitude
and $J-K$ color, and their associated errors, as reported in the 2MASS Point
Source Catalog. The reported photometry is the ``default magnitude'' information
provided in this catalog. Errors in the $J-K$ colors are obtained using the
square root of the quadratic sum of the individual errors reported for the $J$
and $K$ magnitudes. We checked all of the pertinent photometric quality and
contamination flags listed in the 2MASS Point Source Catalog for potentially
problematic stars, and eliminated those stars from further consideration.

Reddening estimates for these stars was obtained from application of the maps of
Schlegel, Finkbeiner, \& Davis (1998), which have superior spatial resolution and
are thought to have a better-determined zero point than the Burstein \& Heiles
(1982) maps.  In cases where the Schlegel et al. estimate exceeds
$E(B-V)_S$ = 0.10, we follow the procedures outlined by Bonifacio, Monai, \&
Beers (2000) to reduce these estimates by 35\%, and obtain an adopted
estimate of reddening, $E\;(B-V)_A$.

In the case of a few of the brighter (nearby) calibration stars, the suggested
reddening was still apparently too high, hence we have assumed zero reddening in
these instances. The final adopted reddening is listed in
column (6) of Table 1. Columns (7) and (8) list the de-reddened $(B-V)_0$ and
$(J-K)_0$ colors, respectively, where we have adopted $E(J-K) = 0.56\; E(B-V)_A$.

\subsection{Spectroscopic Observations}

Medium-resolution spectroscopic data for many of the calibration stars in the
present paper were obtained with the KPNO 2.1m, the Isaac Newton 2.5m on La
Palma, the Las Campanas 2.5m, and the Palomar 5m telescopes. The ESO 1.5m
telescope was also used to obtain observations for a number of the calibration
stars discussed herein, as was the Siding Spring Observatory 2.3m telescope. A
small number of additional observations were obtained with the Kitt Peak
National Observatory 4m telescope, the CTIO 4m telescope, the Anglo-Australian
3.9m telescope, and the McDonald Observatory 2.7m telescope.

A complete discussion of the observing techniques and data reduction procedures
that were followed is presented in Beers et al. (1999), to which we refer the
interested reader. The spectra employed cover (at least) the wavelength region
$\lambda\lambda 3700 - 4500$ {\AA}, at resolutions between 1 and 2 {\AA} (2.5
pixels). Signal-to-noise ratios (at 4000 {\AA}) for the spectra varied from a
minimum of 10/1 to greater than 30/1, with a typical value on the order of 20/1
per pixel. Sample spectra for several of the calibration stars, covering a range
of T$_{\rm eff}$ and [C/Fe], are shown in Figure 1. Since the calibration spectra
were taken as part of observing runs dedicated to survey efforts, no flux
calibrations are available.  

\begin{figure*}[h]
\begin{center}
\includegraphics[angle=270,scale=.7]{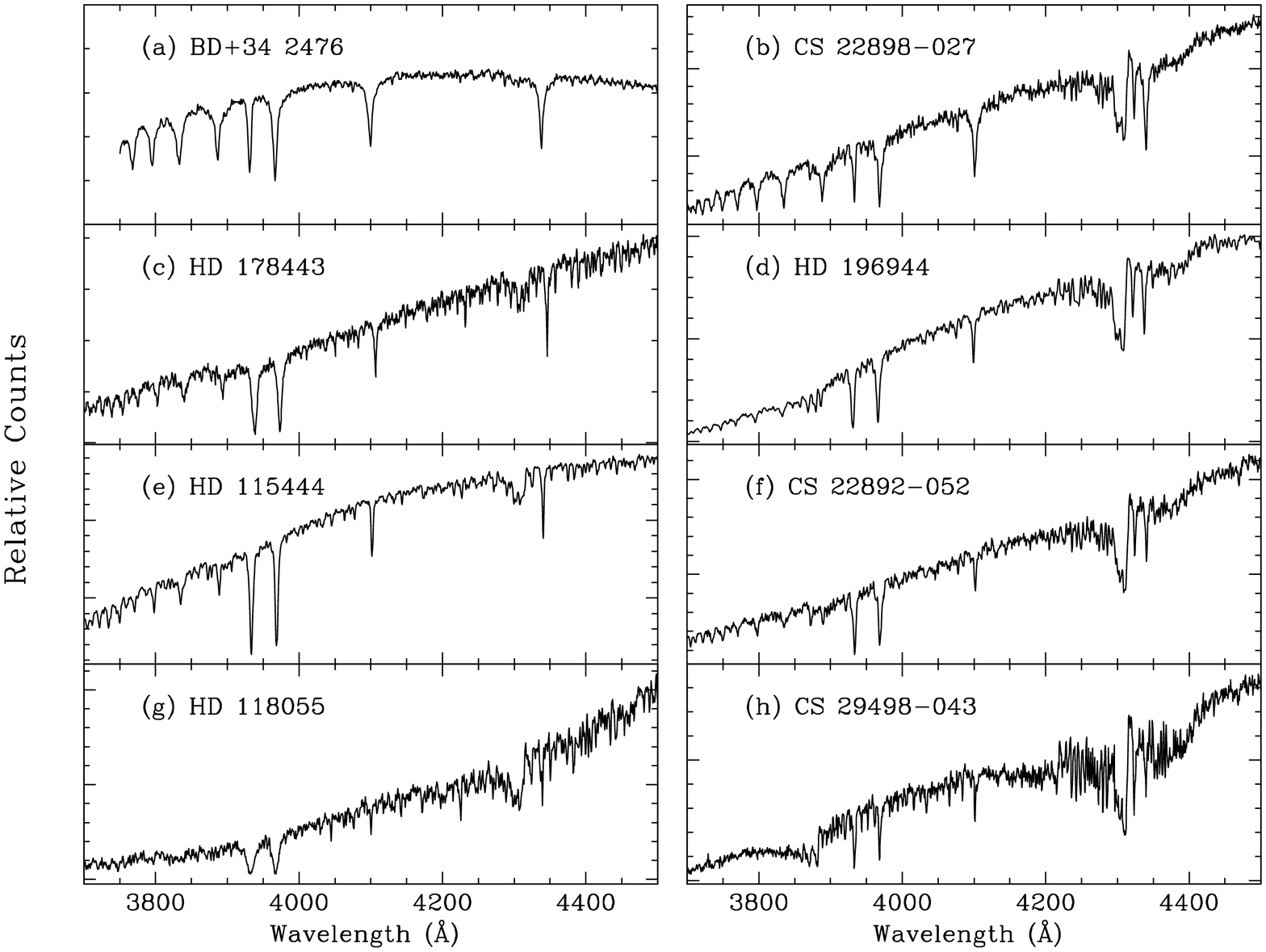}
\end{center}
\caption{Example medium-resolution spectra for calibration stars
covering a range of T$_{\rm eff}$ and [C/Fe]. The warmer stars are at the top;
the cooler stars are at the bottom.  The left-hand panels show stars
with low values of [C/Fe], while those on the right-hand panels show stars with moderate to 
high abundances of [C/Fe]}
\label{} 
\end{figure*}

For the present calibration we require measurements of the line indices $KP$,
a (band-switched) pseudo-equivalent width measure of the CaII K line,
and $GP$, a fixed 15-\AA\ wide band measure of the strength of the CH
G-band. The linebands and sidebands for these two estimators are listed in Table
2. Repeat observations of these stars indicates that the determination of an
individual line index is generally accurate to on the order of $\sim 10$\%. 

Figure 2 indicates, for two CEMP stars of moderate and high [C/Fe] values, the
location of the linebands and sidebands for the $KP$ and $GP$ indices. The
method of estimating the pseudo-equivalent widths on which these indices are
based employs a linear estimation of the local continuum between the sidebands.
As can be appreciated from inspection of this Figure, we do not expect any
difficulties with the estimate of $KP$, except perhaps in cases where extremely
carbon-rich stars might compromise the sidebands of this index. By calibrating
estimates of [Fe/H] directly from stars that exhibit a wide range of carbon
enhancements, as discussed below, we expect to be fairly insensitive to changes
in $KP$ that might be introduced, e.g., by strong molecular carbon features which
can occur in the region of CaII K. By way of contrast, comparison of panels (b) and
(d) of Figure 2, for CS~22892-052 and CS~29498-043, respectively, indicates that
there are certainly concerns raised in the choice of the continuum for the more
carbon-enhanced star, CS~29498-043. In the case of this star, both the blue and
the red sideband are buried in absorption from molecular carbon features, which
may lead to difficulties with the determination of the $GP$ index. Inclusion of
such extreme stars in our calibration set may partially relieve this problem,
but caution should be employed for application of our line-index techniques for
similar extreme stars.   

\begin{figure*}[h]
\begin{center}
\includegraphics[angle=270,scale=.7]{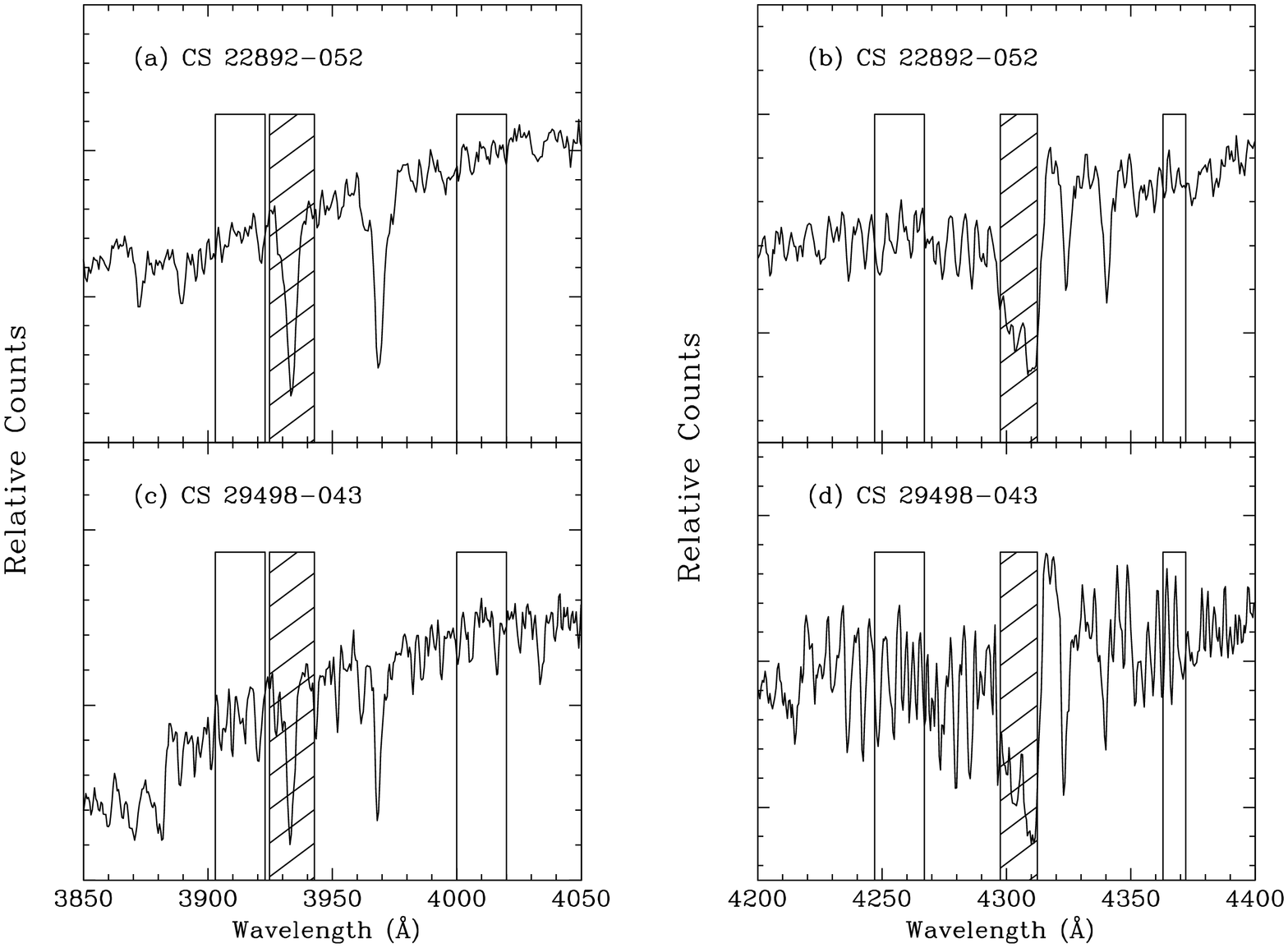}
\end{center}
\caption{Location of the line indices $KP$ and $GP$ for two CEMP stars
of moderate (CS~22892-052) and high (CS~29498-043) carbon enhancement,
respectively. The open rectangles indicate the range over which the sidebands
are estimated. The hatched rectangles indicate the range over which the
linebands are estimated. Note that while one might expect reasonable [C/Fe]
results for stars like CS~22892-052, caution must be employed for stars like
CS~29498-043, where the sidebands are buried in molecular carbon features. }
\label{} 
\end{figure*}

The spectroscopic observations of the calibration stars are summarized in Table
3. Column (1) lists the star name. In column (2) we list the telescopes involved
with the acquisition of the medium-resolution spectra, using codes defined in
the table end notes. The $KP$ and $GP$ line indices are listed in columns (3) and
(4), respectively. The remaining columns in this table are described below.

\section{Calibration of [Fe/H] and [C/Fe]}

We have investigated three alternative techniques for obtaining carbon abundance
estimates of metal-deficient stars, an ANN approach, a regression approach, 
and spectral synthesis. We describe the calibrations of these approaches separately below.  

For all of these exercises we require independent knowledge of [Fe/H] and
[C/Fe], which we obtain primarily from the recent literature. In a few cases, in
particular for the more metal-rich stars, we have had to ``reach back'' to
include a few classic results based on older data. Table 3 lists the pertinent
information for each star. In columns (5)-(7) we list the adopted [Fe/H], [C/H],
and [C/Fe], respectively. The sources used in order to obtain these numbers
(generally from an average of the reported values) are listed in column (8). For
a few stars new analyses appeared quite recently; in some (but not all) cases,
we altered our adopted values to take this new information into account. For the
spectral synthesis method described below, it is also necessary to adopt values
of the physical parameters of the pertinent stellar models used. We obtain these
based on average values of the parameters T$_{\rm eff}$, log g, [Fe/H], and
$\xi$ from the listed sources, and present these values in Column (9). Note
that, for consistency of the spectral synthesis modeling with the predicted
[C/Fe] for the BPSII ``strong G-Band'' stars discussed below, we employed an
input microturbulance value of $\xi = 2.0$ km/sec for estimation of the
[C/Fe]$_{\rm S}$ values, as discussed below.

\subsection{Training of an Artificial Neural Network and a Regression
Alternative}

Artificial Neural Networks are playing a growing role in the analysis of
astronomical data, as they have a number of advantages over more traditional
approaches to data analysis (for a more complete discussion of applications of
ANNs to spectroscopic analyses, see Snider et al. 2001). Here we present a
particularly straightforward application. We seek to estimate [Fe/H] from the
$KP$ indices and $(J-K)_0$, and [C/Fe] based on two alternative sets of inputs:
(1) $KP$, $GP$, and $(J-K)_0$, and (2) from the $KP$ and $GP$ indices alone. In
both cases, we actually make use of the log$_{10}$ of the $KP$ and $GP$ indices, so
that they are on the same logarithmically varying scale as the photometric
inputs and the output quantities.  

We have also explored the use of a multiple regression approach, similar to that
used by Beers et al. (1999) for the prediction of [Fe/H] based on the $KP$ index
and $(B-V)_0$ color, in order to obtain [Fe/H] and [C/Fe] (although here we
employ $(J-K)_0$ colors). We prefer the ANN approach for two reasons. First, one
can train a suitable network very quickly, which allows considerable flexibility
in testing for the presence of potential outliers that could influence the final
results. Secondly, the ANN approach allows for non-linear interactions of the
predictor variables over the parameter space, as well as for {\it different}
interactions of the predictor variables over the calibration space, whereas a
traditional regression approach forces a particular regression model to apply
over the entirety of the calibration space. As a final justification, as new (or
improved) calibration data become available, the ANN approach allows for rapid
retraining. Many users may be more familiar with the usual regression
techniques, hence we have made available approximate regression relationships
for our predictions that reasonably reproduce the results of the ANN approaches.

A first-pass training of the two sets of ANNs was accomplished using the inputs
given above, and six hidden nodes, in order to predict the desired outputs
[Fe/H] and [C/Fe]. Although it is possible to simultaneously predict the two
output variables, our experiments suggested that (slightly) better results were
obtained using dedicated networks for each output.  The networks trained to
predict single variables exhibited scatter in the derived abundances
that were typically 10-20\% lower than the networks that predicted the two
variables at once.  In order to differentiate the
various networks we define them as follows:

\medskip

$${\rm [Fe/H]}_{\rm A} = f\;[LKP,(J-K)_0]$$

$${\rm [C/Fe]}_{\rm A} = f\;[LKP,LGP)]$$

\medskip

\noindent In the above expressions $LKP$ and $LGP$ indicate the log$_{10}$ of the $KP$
and $GP$ indices, respectively. We also trained a network for estimation of
[C/Fe] that took advantage of using available $(J-K)_0$ photometry as an
additional input, but this produced very little improvement in the residuals for
[C/Fe], hence we decided to proceed with the generally more applicable estimate
that does not require photometry as an input.

For calibration of these networks, 60\% of the data is used as a training
subsample, while 20\% of the data is set aside for a testing subsample. The
remaining 20\% of the data, which is never seen by the ANNs, is used as a
validation set in order to obtain an independent indication of the accuracy with which
the desired outputs can be predicted.  We also tested the stability of the
networks by drawing ten different sets of training, testing, and validation
subsets, and looking for any indication of inconsistency in the predictions of
the output quantities.  None were found.

During the course of our training of the ANNs, it was immediately noticed that
the ANNs had difficulty in obtaining reasonable estimates of [Fe/H] for stars with
[Fe/H] $> -1.0$, a problem that is almost certainly due to saturation, or near
saturation, of the $KP$ index for metal-rich and/or very cool stars. Similar
problems were encountered in the Beers et al. (1999) calibration, but were
addressed by introduction of a second estimator, the Auto-Correlation Function
(ACF). A similar adjustment is made difficult for carbon-rich stars, because the
molecular carbon bands themselves can have a large effect on an ACF, at least as
defined in the Beers et al. (1999) paper.

Even after removal of stars that were expected to present problems because of
possible saturation effects, it was clear that a number of stars in the
calibration set exhibited rather large deviations, so the training was repeated
with these stars set aside. Such ``problem'' stars might be also indicative of
poorly measured $KP$, $GP$, or $(J-K)_0$, or errors in the high-resolution
estimates of [C/Fe] and/or [Fe/H] reported for the calibration stars themselves.
Our final results are summarized in Table 4. Column (1) of Table 4 reports the
star names, column (2) lists the adopted external estimate of [Fe/H] from the
literature, column (3) lists the estimate of [Fe/H] obtained from the ANN
analysis, [Fe/H]$_{\rm A}$, along with its associated residual relative to the
literature value of [Fe/H], and column (4) lists the estimate of [Fe/H],
[Fe/H]$_{\rm R}$, obtained from the regression analysis, described below, along
with its associated residual. Column (5) lists the adopted literature value of
[C/Fe], column (6) lists the estimate of [C/Fe] obtained from the ANN analysis,
[C/Fe]$_{\rm A}$, along with its associated residual relative to the literature
value of [C/Fe], and column (7) lists the estimate of [C/Fe] obtained from the
regression analysis, [C/Fe]$_{\rm R}$, described below, along with its
associated residual. Column (8) indicates the grouping of the calibration stars into
several bins in T$_{\rm eff}$, [Fe/H], and [C/Fe], respectively, as listed in
the table end notes. We use these bins below to explore the typical
residuals from the two methods used to estimate [Fe/H], and the three methods
used to estimate [C/Fe] over the calibration space. Column (9) lists the
spectral synthesis estimate of [C/Fe], [C/Fe]$_{\rm S}$, as described below, and
its residual with respect to the literature value of [C/Fe]. Column (10)
indicates, with an ``X'', the stars which were dropped from the [C/Fe]
calibration due to their large residuals in estimation of [C/Fe] by the ANN
approach (dropped stars are those with an absolute [C/Fe]$_{\rm A}$ residual
larger than 0.6 dex).  

Figures 3(a) and 3(b) show the distribution of [C/Fe] and [C/H], respectively, as a
function of [Fe/H], for the stars that were used for our final calibration
exercise.   Figures 3(c) and 3(d) show the distribution of these same quantities as a
function of T$_{\rm eff}$.  As can be appreciated from inspection of these
figures, it would be desirable to obtain additional observations of well-studied 
stars with [C/Fe] $> +1.0$ for metallicities [Fe/H] $> -2.0$.  Similarly, we
would benefit from observations of additional stars with Teff $> 5500$~K. 

\begin{figure*}[h]
\begin{center}
\includegraphics[angle=270,scale=.7]{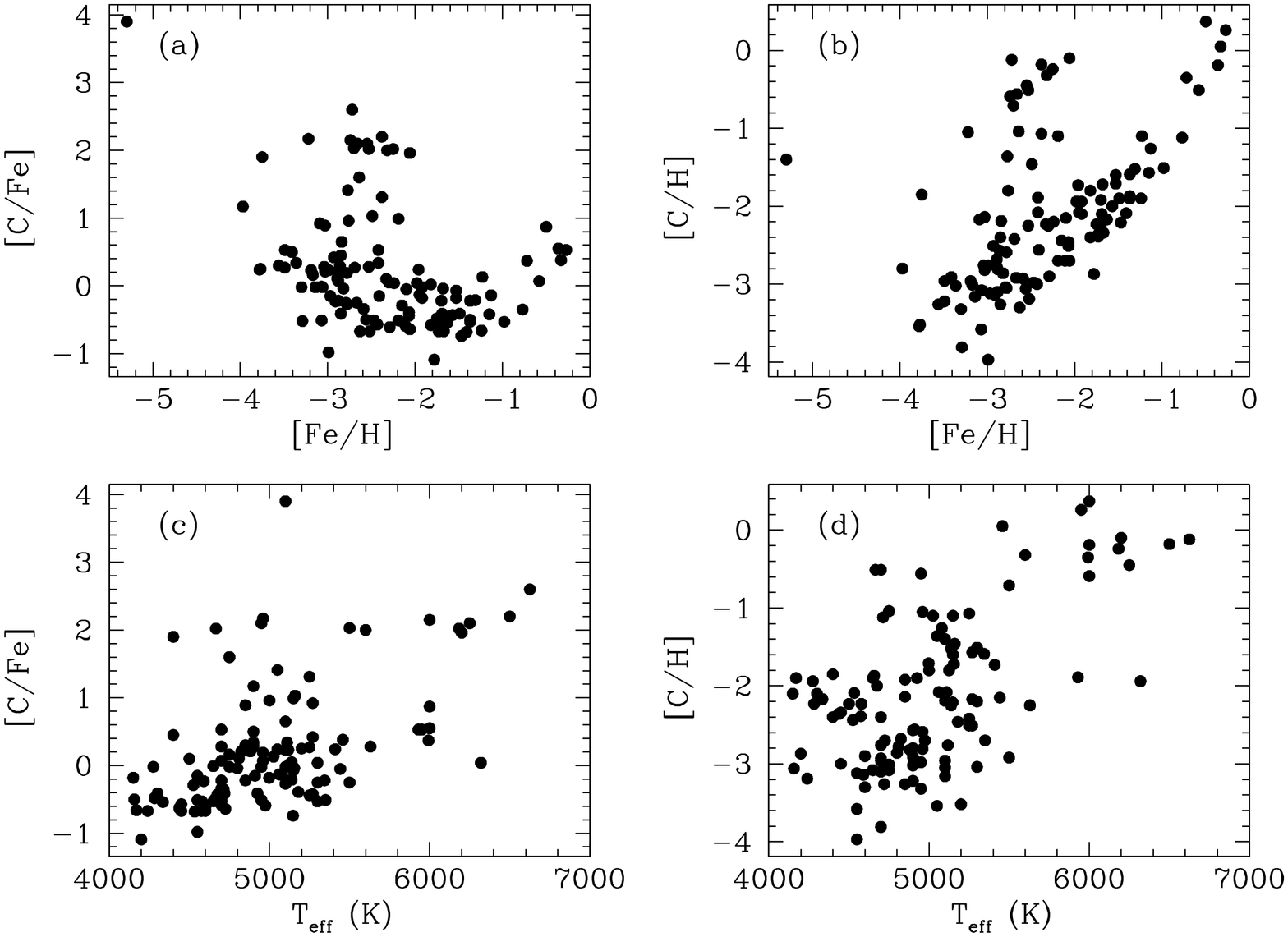}
\end{center}
\caption{Distribution of adopted [C/Fe] and [C/H] from the recent
literature for the accepted calibration stars as functions of [Fe/H] (panels a
and b), and as a function of Teff (panels c and d).
}
\label{} 
\end{figure*}

Figure 4 shows the distribution of positive and negative residuals, with the
size of the points scaled to be proportional to the size of the individual
residual, for the ANN determinations of [Fe/H] and [C/Fe] with respect to the
adopted literature values, as functions of the predictor variables used in the
estimates. Inspection of the patterns and sizes of the residuals indicates that,
at least over the calibration space we have considered, the residuals are evenly
distributed in their sizes and signs. More quantitative estimates of the
residuals are presented in Table 5, discussed below.

\begin{figure*}[h]
\begin{center}
\plotone{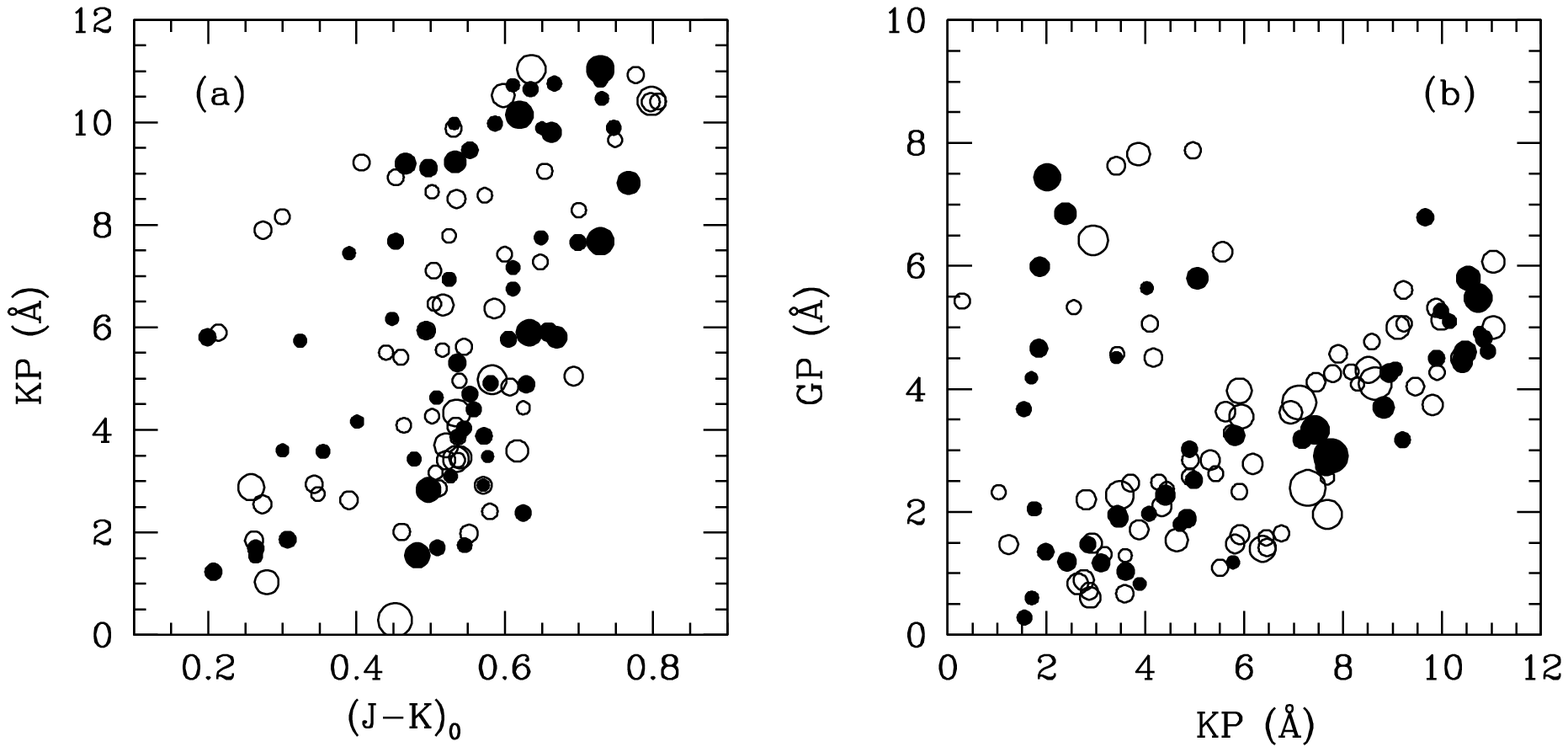}
\end{center}
\caption{(a) Distributions of positive and negative residuals of the
ANN estimates of [Fe/H], with respect to adopted literature values, for the
accepted calibration stars over the calibration space defined by $KP$ and $(J-K)
_0$.  (b) Distribution of positive and negative residuals of the ANN estimates
of [C/Fe], with respect to the adopted literature values over the calibration
space defined by $GP$ and $KP$.  In both cases, positive residuals are indicated
by filled circles, while negative residuals are indicated by open circles. 
The size of each point is scaled to be proportional to the size of the residual.  
}
\label{} 
\end{figure*}

Approximate regression relations for [Fe/H] and [C/Fe] are:
\medskip

\begin{eqnarray}
{\rm [Fe/H]}_{\rm R} = -1.272\; (0.196) - 1.562\; (0.588)\; LKP \nonumber \\
+ 3.906\; (0.411)\; LKP^2 - 4.045\; (0.203)\; (J-K)_0 \nonumber \\ 
\epsilon = 0.25\;, \; {\rm R}^2 = 0.95 \nonumber \\
{\rm [C/Fe]}_{\rm R} = 1.698\; (0.071) - 3.403\; (0.108)\; LKP \nonumber \\
+ 1.995\; (0.090)\; LGP \nonumber \\ 
\epsilon = 0.26\; , \;{\rm R}^2 = 0.95 \nonumber \\
\end{eqnarray}
\medskip

\noindent In the above, $\epsilon $ indicates the expected prediction error of the
estimate, while R$^2$ is the total variance that can be accounted for by the
regression relationship. The quantities in parentheses following each of the 
coefficients are their one-sigma errors.  As in the case of the ANNs, a
regression was performed that predicts [C/Fe] based on the inclusion of 
$(J-K)_0$, in addition to $LKP$ and $LGP$, but little improvement over the
regression that did not involve photometry was obtained.

\subsection{The Spectral Synthesis Approach}

The first stage in the synthesis calculations requires that reasonable
first guesses be made for the pertinent physical parameters of the stellar atmospheres
that are employed. In the case of the calibration stars, these were already
known based on previous high-resolution studies.  For the program stars, the
``strong G-band'' stars of BPSII, we employed estimates of [Fe/H] derived
from the ANNs, and approximations of T$_{\rm eff}$ and log g, as described below.

We computed synthetic spectra of calibration and program stars over essentially
the entire wavelength range of the CH G-band, 4190~\AA~$\leq$~$\lambda$~$\leq$~4425~\AA. 
The current version of the LTE line analysis code MOOG (Sneden 1973) was used to
generate these spectra. Stellar atmospheres were interpolated from the grid of
ATLAS9 models by Castelli \& Kurucz (2003), using software developed by A.
McWilliam and I. I. Ivans (private communication). The values of T$_{\rm eff}$,
log g, and [M/H] that were adopted for each star are provided in column (9) of
Table~3 (calibration stars) and Table~7 (program stars). We culled Kurucz's
atomic and molecular line database\footnote{http:
//kurucz.harvard.edu/linelists.html} to produce the input line list of more than
3000 neutral, singly ionized, and CH molecular species. We also assumed ${\rm
^{12}C/^{13}C}$~=~10, but this parameter also did not strongly influence the
derived C abundances. We did not include the CN lines that exist in the
4190--4215~\AA\ region, after preliminary synthetic spectrum experiments showed
that CN absorption produced almost no effect on the derived C abundances, which
mostly were estimated from the 4230--4370~\AA\ region.

In our computations we assumed solar abundances in good agreement with those
recommended by Lodders (2003). We also adopted log~$\epsilon$(C)$_\odot$~= 8.56
and log~$\epsilon$(O)$_\odot$~= 8.92, which were generally employed prior to
their re-examination by Allende Prieto et al. (2001,2002) and Holweger (2001).
The best values of these two abundances now appear to be approximately
log~$\epsilon$(C)~= 8.4 and log~$\epsilon$(O)~= 8.7, about 0.2~dex lower.
Adoption of these new photospheric values would lead only to a uniform offset
from our quoted C abundances.  

Some caution should be observed, however, in interpreting our abundances for
the coolest, most metal-rich stars, for which CO formation potentially could
have a larger influence on the derived C abundance.
Relative [X/Fe] abundances of the $\alpha$-elements O, Mg, Si, and Ca were 
set to typical values for halo stars, [$\alpha$-element/Fe]~=~+0.4. 
Finally, we assumed a uniform value for microturbulent velocity, 
v$_{\rm t}$~=~2~km~s$^{\rm -1}$ (tests conducted with $\pm$0.5 variations 
in this parameter demonstrated the relative insensitivity of
derived C abundance to v$_{\rm t}$). 
Our tests indicated that small changes in
carbon isotopic ratios, $\alpha$-element enhancements, and microturbulance
produced only small variations ($\lesssim$0.1~dex) in our estimated carbon
abundances.

The synthetic spectra were computed for different values of [C/Fe], and smoothed
with a Gaussian function to match the resolution of the observed spectra.
Setting the continuum levels in the observed spectra could not be reliably done
from inspection of the spectra themselves. The data were obtained with many
different several telescope/instrument configurations, all without the benefit
of flux calibration, hence the continuum shape is determined by the convolution
of the actual stellar continuum with the instrumental response function. The
spectra are relatively low resolution for abundance work; in cases of strong
CH, CN, and/or C$_{\rm 2}$ bands true continuum regions are difficult to
identify. However, for each instrumental setup, at least a few warm, very
metal-poor, very weak-CH stars were observed. We used specialized software
(Fitzpatrick \& Sneden 1987) to co-add these weak-lined spectra and then
interactively fit spline functions through their easily identified continuum
points spaced at roughly 50~\AA\ intervals. The resulting approximate continuum
curves were divided into the spectra of all of the calibration and program
stars.

After the spectra were flattened, the computed and observed spectra were
visually compared; difference plots between the two were also compared. These
led to carbon abundances estimated to the nearest 0.05~dex, for stars with high
signal-to-noise spectra and moderate-to-strong CH band strengths. For stars with
barely detectable CH features and/or noisy spectra, the errors on the abundance
estimates are larger. The results of the synthesis calculations of [C/Fe] for
the calibration stars are provided in column (9) of Table~4, along with the
residual obtained from a comparison with the results from the high-resolution
studies. For moderate to higher signal-to-noise spectra, the typical error in
estimation of [C/Fe] by the spectrum synthesis technique, 
as judged from the residuals with respect to the external adopted values of
[C/Fe], is on the order of 0.20-0.25 dex. In column (9) of Table~4 we note with
single colons those abundances that are likely to have larger uncertainties, as judged
by the quality of the fit to the observed spectra.  Double colons indicate stars
with very poor fits, which should be considered quite uncertain.   

Figure 5 shows several examples of the synthetic/observed spectrum
matches for the calibration stars. In this Figure, we show two examples for each
of our four bins in T$_{\rm eff}$, one each with low [C/Fe] ([C/Fe]$_{\rm S} \le 0.0$),
and one each with large [C/Fe] ([C/Fe]$_{\rm S} > +1.0$).  These are the same
stars for which the entire available spectra are shown in Figure 1.

\begin{figure*}[t]
\begin{center}
\epsscale{.8}
\plotone{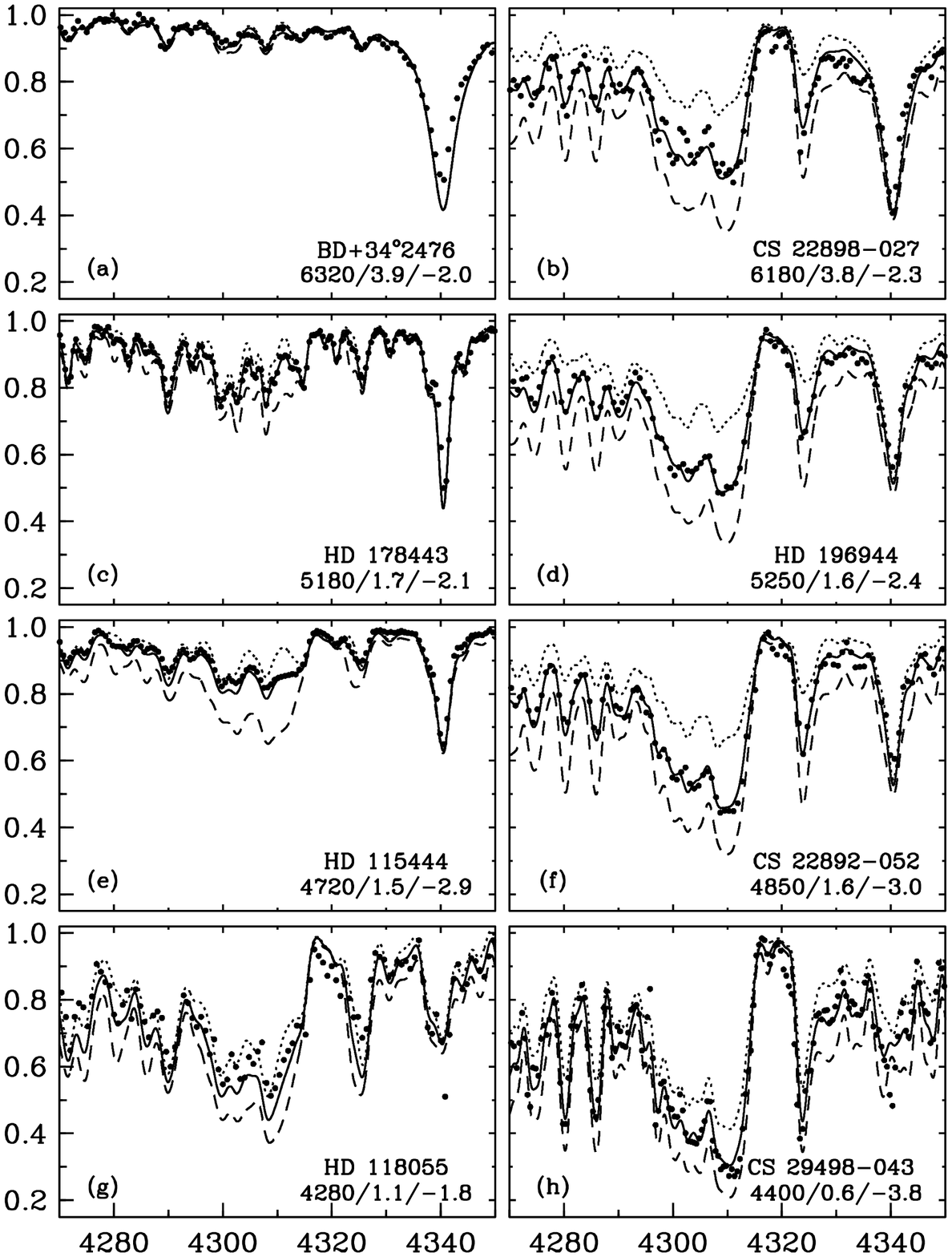}
\epsscale{1}
\end{center}
\caption{Examples of the spectral synthesis fits for calibration
stars covering a range of T$_{\rm eff}$ and [C/Fe]. The warmer stars are at the
top; the cooler stars are at the bottom. The left-hand panels show stars with
low values of [C/Fe], while those on the right-hand panels show stars with
moderate to high abundances of [C/Fe]. The values of the adopted atmospheric
model parameters are indicated (T$_{\rm eff}$/log g/ [M/H]). The data are shown
as filled circles. The lines indicate the fits obtained from various input
values of [C/Fe] differing by 0.5 dex; the adopted best-fit is shown as a bold line. }
\label{} 
\end{figure*}

As noted above, our calibration sample is unavoidably weighted toward the
inclusion of subgiants and giants, so the effect of surface gravity on our
estimation of [C/Fe] could not be adequately determined from this sample alone. 
Instead, we explored this issue by synthesizing spectra of higher gravity
stars, up to log g $= +5.0$, and applying our present techniques to their
analysis.  These experiments indicated that, over our calibration space, the
effect of higher surface gravity introduced shifts of from a few tenths of
a dex to (in the most extreme cases) 0.5 dex in the determination of [C/Fe]. 

\subsection{Comparison of the Various Methods}

In this section we consider the accuracies of the methods we employ in more
detail. First, we examine plots of the residuals of the predicted [Fe/H] and
[C/Fe] for the three methods as functions of their individual predictor
variables and as compared to our adopted literature values of these quantities.
We then quantitatively consider the accuracies of the methods we have explored
for estimation of [Fe/H] and [C/Fe], first globally, and then across the various
bins in T$_{\rm eff}$, [Fe/H], and [C/Fe] as defined in the end notes of Table
4. 

\begin{figure*}[h]
\begin{center}
\plotone{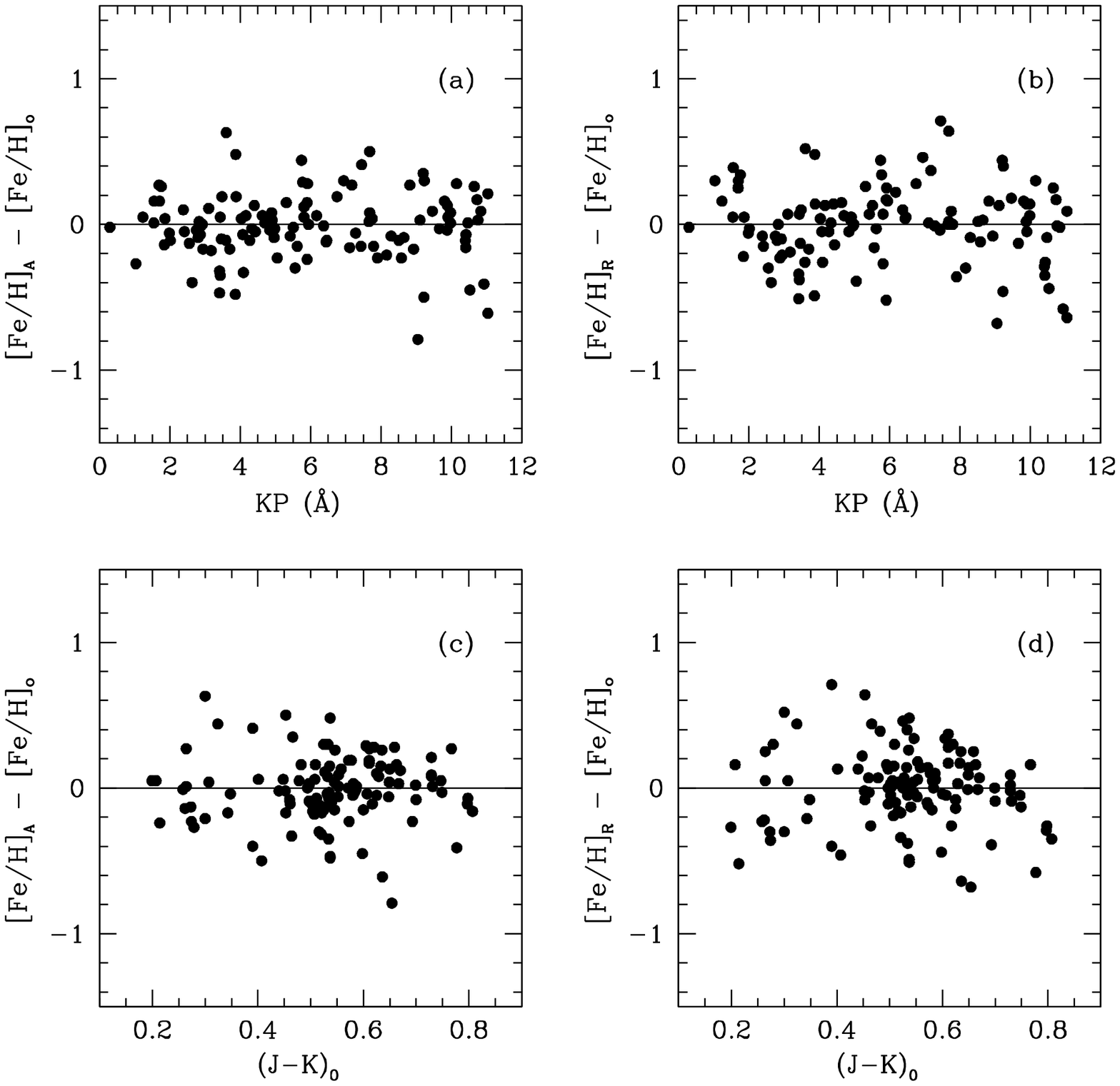}
\end{center}
\caption{(a) Distributions of residuals of the estimated [Fe/H] based on the
ANN approach, [Fe/H]$_{\rm A}$, as compared to the adopted literature value,
[Fe/H]${\rm o}$, as a function of the predictor variable $KP$.  (b) The same as
in panel (a), but for [Fe/H]$_{\rm R}$. (c) Distributions of residuals of the
[Fe/H]$_{\rm A}$ as compared to [Fe/H]${\rm o}$, as a function of the predictor
variable $(J-K)_0$. (d) The same as in panel (c), but for [Fe/H]$_{\rm R}$.   
}
\label{} 
\end{figure*}

\begin{figure*}[h]
\begin{center}
\plotone{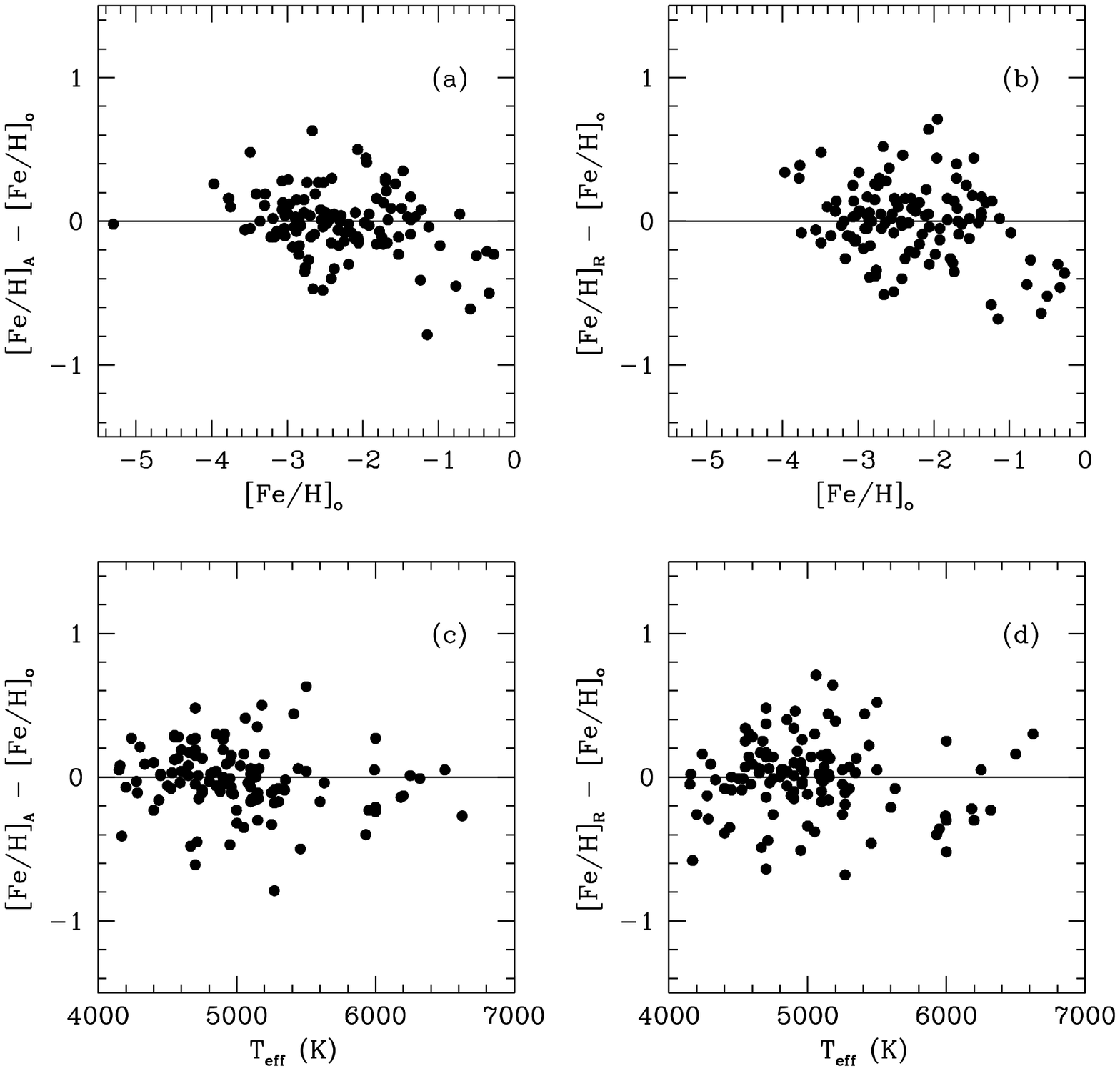}
\end{center}
\caption{(a) Distributions of residuals of the estimated [Fe/H] based on the
ANN approach, [Fe/H]$_{\rm A}$, as compared to the adopted literature value,
[Fe/H]$_{\rm o}$, as a function of [Fe/H]$_{\rm o}$.  (b) The same as
in panel (a), but for [Fe/H]$_{\rm R}$. Note that the residual for HE~0107-5240
is not shown (see text).  (c) Distributions of residuals of the
[Fe/H]$_{\rm A}$ as compared to [Fe/H]${\rm o}$, as a function of T$_{\rm eff}$.
(d) The same as in panel (c), but for [Fe/H]$_{\rm R}$. 
}
\label{} 
\end{figure*}

Figure 6 shows the residuals in the determination of the estimated [Fe/H], as
compared to the individual predictor variables $KP$ and $(J-K)_0$, for both the
ANN and regression approaches. As can be seen from inspection of this Figure,
the residuals are distributed similarly over the predictor variables for both
techniques. Some of the largest negative residuals in the estimates of [Fe/H]
occur when either $KP$ is large, or for the redder stars; at these extremes the
$KP$ index is approaching saturation. Figure 7 shows the same sets of residuals
as a function of the adopted literature values of [Fe/H] and T$_{\rm eff}$. Note
that the residual for HE~0107-5240 is not plotted in panels (b) and (d), since
it is extremely large; the polynomial used for the estimate of [Fe/H]$_{\rm R}$
for this [Fe/H] $= -5.3$ star provides an unrealistic estimate of its
metallicity. The larger negative residuals for [Fe/H] $> -1$ are clearly evident
in panels (a) and (b). 

\begin{figure*}[h]
\begin{center}
\plotone{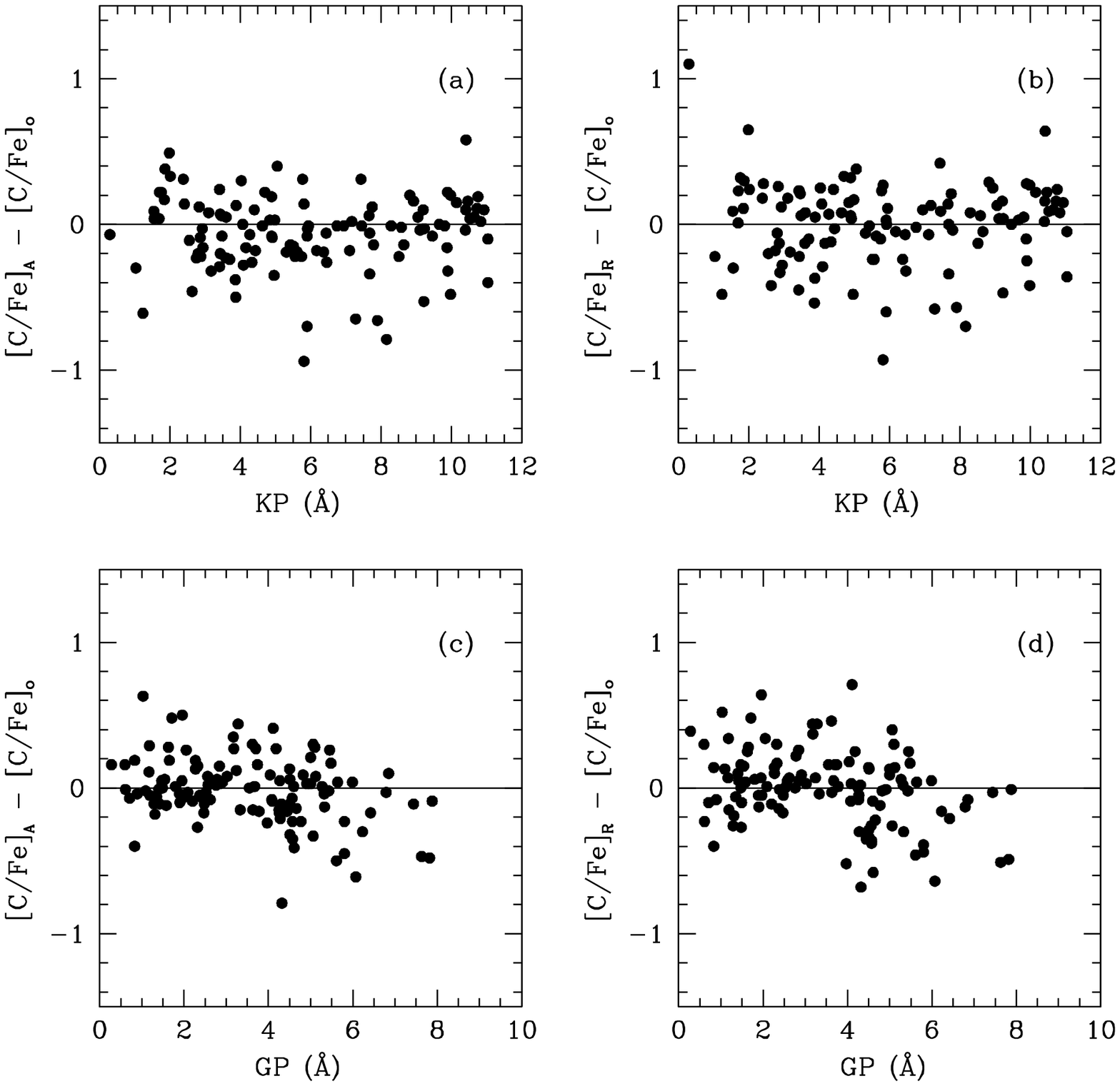}
\end{center}
\caption{(a) Distributions of residuals of the estimated [C/Fe], based on the
ANN approach, [C/Fe]$_{\rm A}$, as compared to the adopted literature value,
[C/Fe]${\rm o}$, as a function of the predictor variable $KP$.  (b) The same as
in panel (a), but for [C/Fe]$_{\rm R}$. (c) Distributions of residuals of the
[C/Fe]$_{\rm A}$ as compared to [C/Fe]${\rm o}$, as a function of the predictor
variable $GP$. (d) The same as in panel (c), but for [C/Fe]$_{\rm R}$.   
}
\label{} 
\end{figure*}

\begin{figure*}[h]
\begin{center}
\plotone{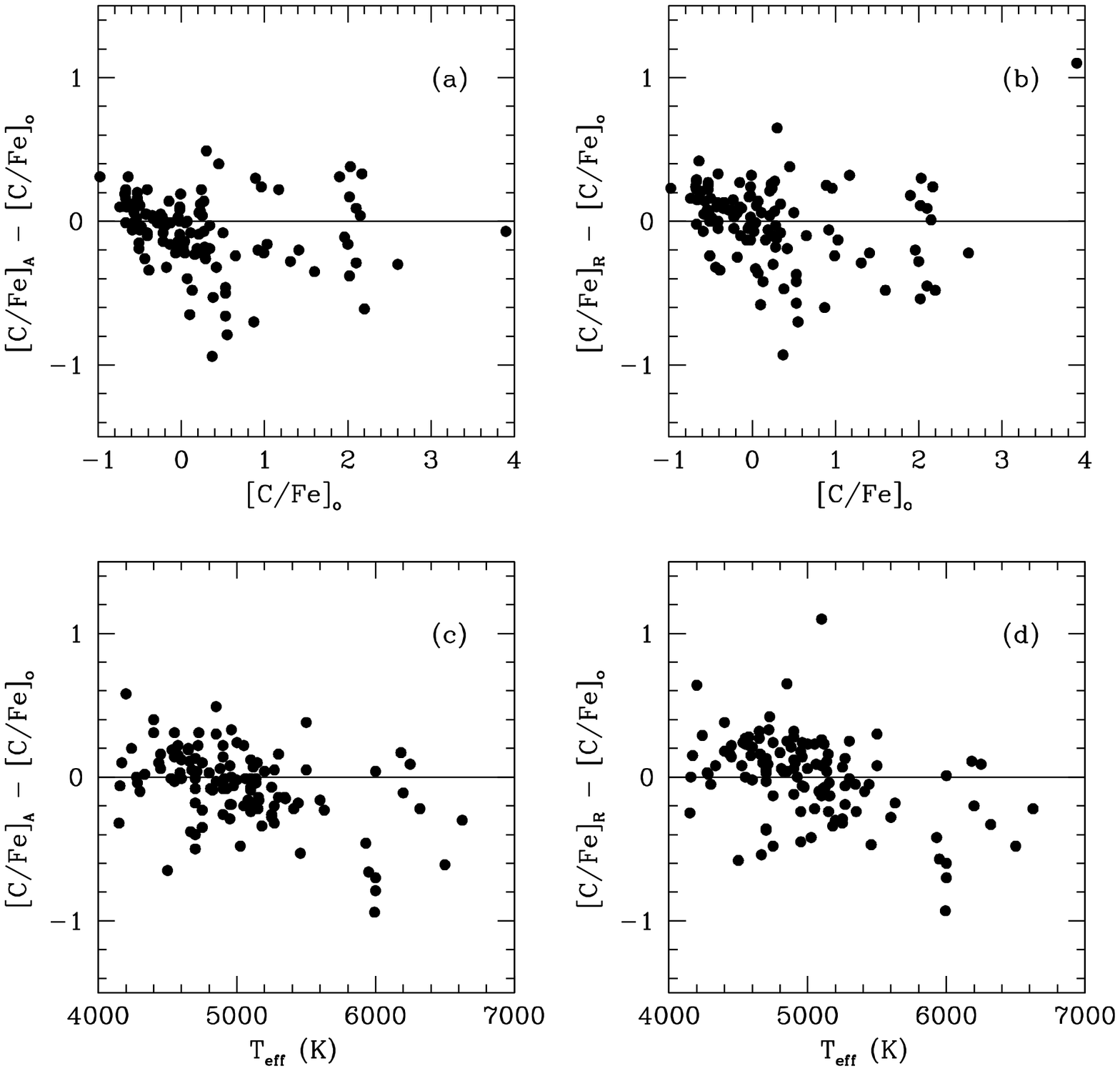}
\end{center}
\caption{(a) Distributions of residuals of the estimated [C/Fe], based on the
ANN approach, [C/Fe]$_{\rm A}$, as compared to the adopted literature value,
[C/Fe]$_{\rm o}$, as a function of [C/Fe]$_{\rm o}$.  (b) The same as
in panel (a), but for [Fe/H]$_{\rm R}$. (c) Distributions of residuals of the
[C/Fe]$_{\rm A}$ as compared to [C/Fe]${\rm o}$, as a function of T$_{\rm eff}$. 
(d) The same as in panel (c), but for [C/Fe]$_{\rm R}$.   
}
\label{} 
\end{figure*}

Figure 8 shows the residuals in the determination of the estimated [C/Fe], as
compared to the individual predictor variables $KP$ and $GP$, for both the ANN
and regression approaches. As can be seen from inspection of this Figure, the
residuals are distributed similarly over the predictor variables for both
techniques. Some of the largest negative residuals in the estimates of [C/Fe]
occur when $KP$ is large, due to the approaching saturation of this index. For
both approaches there is a tendency to underestimate [C/Fe] by about 0.5 dex
for stars with $GP > 6$ \AA , as this index approaches saturation. Figure 9
shows the same sets of residuals as a function of the adopted literature values
of [C/Fe] and T$_{\rm eff}$.  Note that both the ANN and regression methods
exhibit some of their largest negative residuals between $0 < {\rm [C/Fe]} <
+1.0$, although there are many smaller residuals in this range as well. In
the case of the regression technique, the largest positive residual seen in
panels (b) and (d) is associated with HE~0107-5240, where once more the
regression approach is failing at this extreme value of [C/Fe].  There are a
number of large negative residuals seen in panels (c) and (d) at T$_{\rm eff} > 6000$ K,
where the strength of the $GP$ index is rapidly declining due to its sensitivity
to temperature, and hence is subject to more measurement error.

\begin{figure*}[h]
\begin{center}
\includegraphics[angle=0,scale=.8]{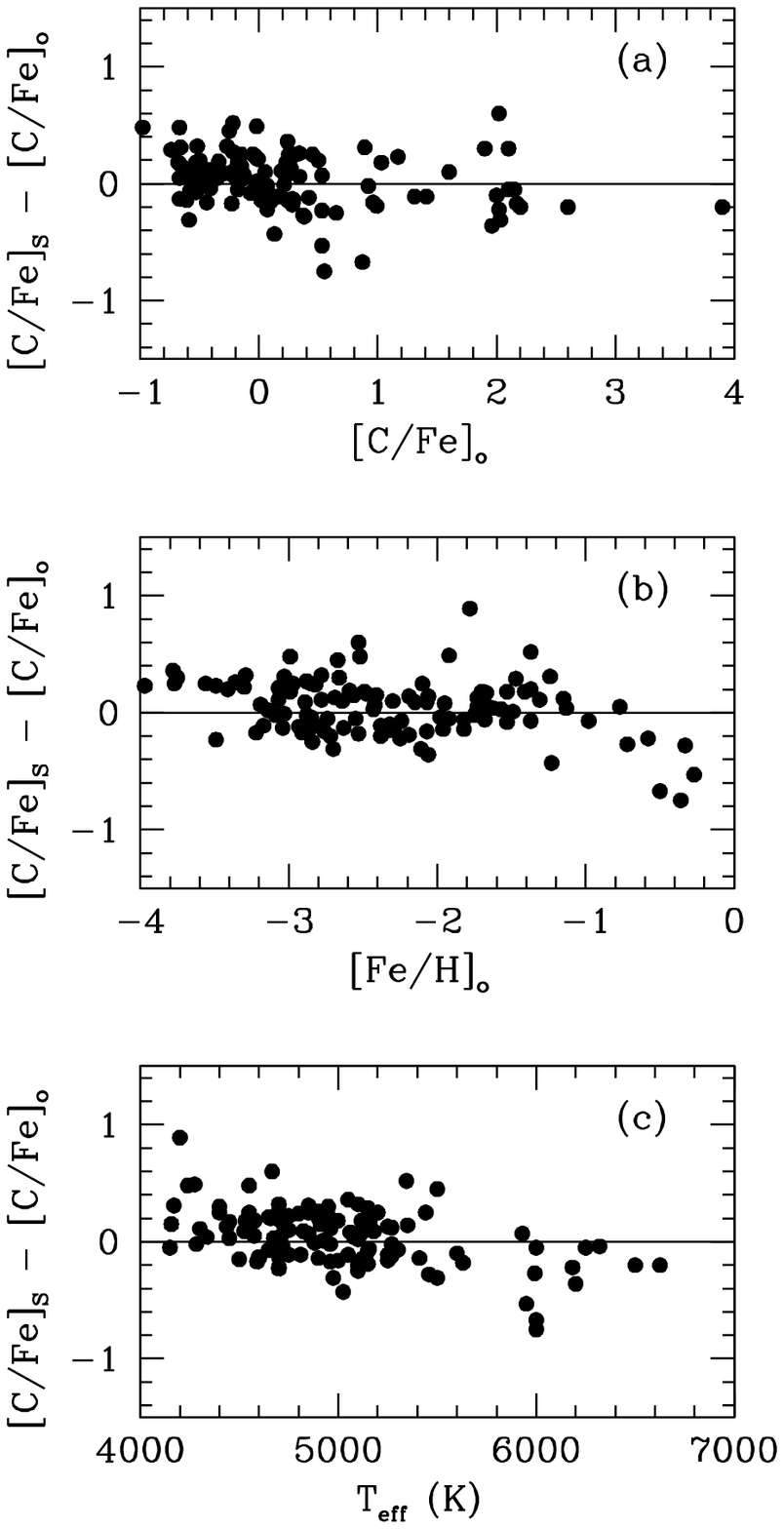}
\end{center}
\caption{(a) Distributions of residuals of the estimated [C/Fe], based on the
synthesis approach, [C/Fe]$_{\rm S}$, as compared to the adopted literature value,
[C/Fe]$_{\rm o}$, as a function of [C/Fe]$_{\rm o}$.  (b) The same as
in panel (a), but as a function of the adopted literature value of [Fe/H]$_{\rm o}$. (c) 
The same as in panel (a), but as a function of T$_{\rm eff}$. 
}
\label{} 
\end{figure*}

Figure 10 shows the residuals in the estimate of [C/Fe] obtained from the spectral
synthesis method as a function of the adopted literature values of [C/Fe],
[Fe/H], and T$_{\rm eff}$, respectively.  Note that some of the largest negative
residuals in panel (a) occur in the same range of [C/Fe]$_{\rm o}$ that proved
problematic for the ANN and regression estimates.  One might be suspicious of
the adopted literature values of [C/Fe] for these stars.  In panel (b) it is
clear that the synthesis estimates of [C/Fe] also appear to exhibit larger
negative residuals at higher metallicity, as was seen for the ANN and regression
estimates. As was also seen previously, in panel (c) a few of the largest
negative residuals occur for T$_{\rm eff} > 6000$, presumably due to the
weakness of the G-band, and the difficulty of matching its strength to the
models when it is quite small.

Figure 11 (a) shows how the ANN estimates of [C/Fe] for the calibration stars
compare with those obtained from the synthesis approach over the entire calibration
space. Figure 11 (b) is the residual of the ANN estimate as compared to the
synthesis estimate, as a function of the synthesis estimate.  With the exception
of a number of individual cases near the solar value, and one star with
[C/Fe]$_{\rm S} \simeq +2.6$, the agreement is quite satisfying, at least given
the very different methodology applied.

\begin{figure*}[h]
\begin{center}
\plotone{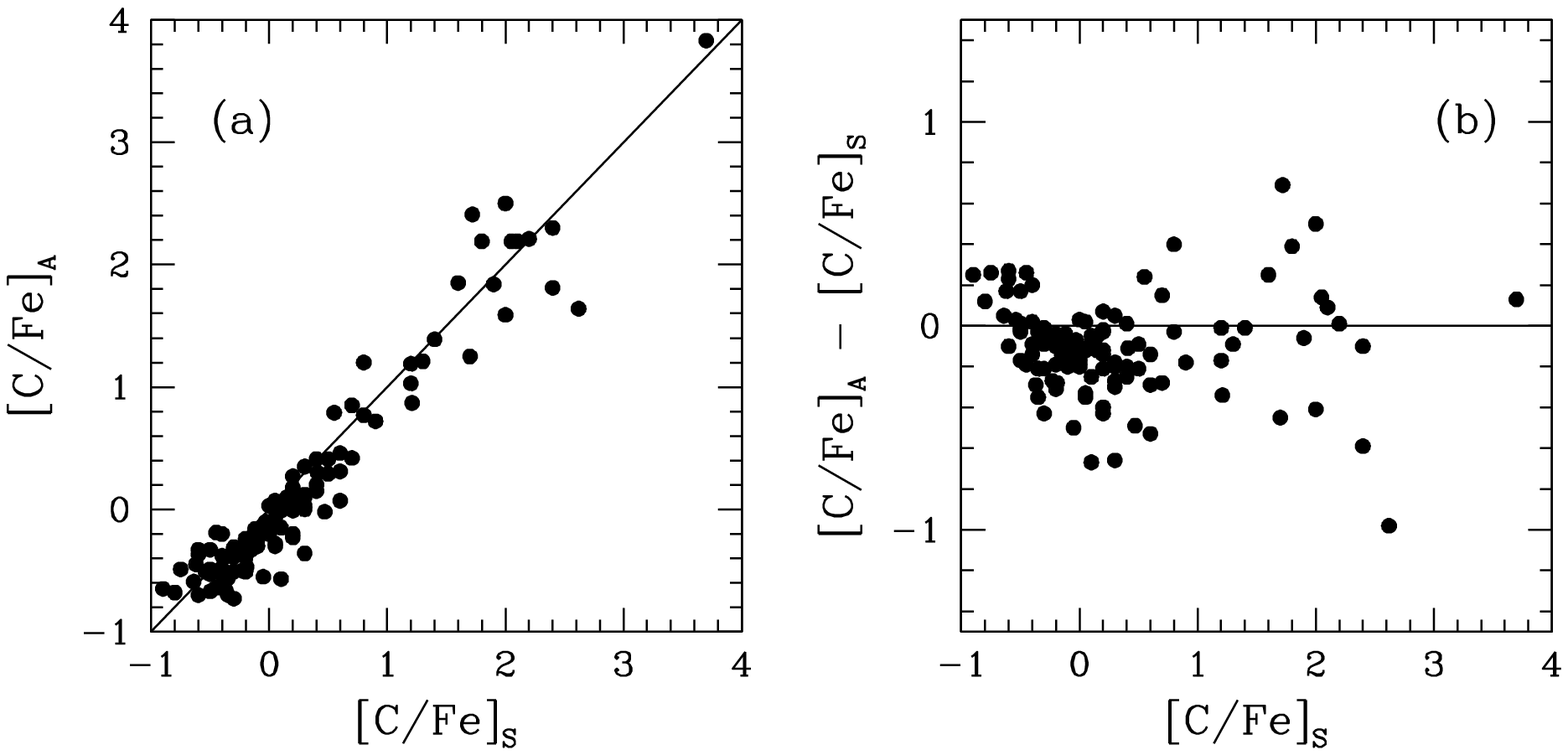}
\end{center}
\caption{(a) Comparison of the estimated [C/Fe] from the ANN
approach, [C/Fe]$_{\rm A}$, with that obtained from the synthesis approach,
[C/Fe]$_{\rm S}$, for the calibration stars. The solid
line is the one-to-one relation. (b) Residuals about the one-to-one line as a
function of [C/Fe]$_{\rm S}$. }
\label{} 
\end{figure*}

We now consider quantitative estimates of the global accuracies obtained, as
compared to the adopted literature values for [Fe/H] and [C/Fe] listed in the
end notes of Table 4. The results are provided in Table 5. Columns (2)-(6)
indicate the mean offset of the estimate under consideration with respect to the
adopted literature values. The quantities in parentheses indicate the standard
deviation of the estimates about the listed mean offset. We have also explored
the accuracy of our derived estimates of [Fe/H] and [C/Fe] over the bins listed
in Table 4. Column (1) of Table 5 lists results for the various bins in T${\rm
eff}$, [Fe/H], and [C/Fe].  

As can be seen from inspection of Table 5, estimates of [Fe/H] from either the
ANN or regression approach exhibit similar offsets and scatter. It should be
noted that the offsets in the estimates of [Fe/H] for the stars with literature
values [Fe/H] $> -1.0$ are low by on the order of 0.3 to 0.4 dex, most likely
due to the saturation or near saturation of the $KP$ index. Similarly, and
presumably for the same reason, the estimates of [C/Fe] are generally offset by
0.4 to 0.5 dex for the ANN and regression approaches for stars with [Fe/H] $>
-1.0$. Interestingly, the synthesis estimates of [C/Fe] suffer from an offset of
0.3 dex for the more metal-rich stars as well.

Other than the obvious problems with the results for the more metal-rich stars,
inspection of Table 5 indicates that the ANN and regression estimates of [C/Fe]
are only slightly worse than obtained from the synthesis results. There are
certainly individual cases, for stars with very strong molecular bands, where
one must be concerned about the results obtained from the line-index approach. For
such stars there is reason to prefer the synthesis approach for estimation of
[C/Fe]. Further discussion of this point is made below.

\section{The ``Strong G-Band'' Stars of Beers, Preston, \& Shectman}
 
The HK objective-prism/interference-filter survey of Beers and colleagues has
discovered numerous carbon-rich stars over a wide range of metal abundance,
extending down to the lowest abundance stars presently known. In a previous
paper, BPSII noted 56 stars from their survey with possibly strong G-bands relative to
other metal-deficient stars of similar colors. These stars, listed in Table 8 of
BPSII, form our program sample. For many of these stars the original spectra that
were available (from the original HK survey observations) were not of uniformly
high quality (and in some cases were limited in the spectral range covered) to
perform a satisfactory analysis, so we obtained additional higher
signal-to-noise spectra for the majority of them, using the same
telescopes and spectrographs that were employed for the calibration objects. A
number of the BPSII ``strong G-band stars'' have been observed subsequently at
higher spectral resolution (indeed, several appear among our calibration
objects), which provides an additional check on our ability to estimate
metallicities and carbon abundances from medium-resolution spectra of CEMP stars.

Photometric information for our program stars is supplied in Table 6. The column
definitions, as well as procedures used to obtain the information listed, are identical
to those described for the calibration stars listed in Table 1.  One star,
CS~30493-064, was not found in the 2MASS catalog, hence we have no $(J-K)_0$ for
it, and therefore cannot obtain estimates of [Fe/H] for this object.

\subsection{Determination of [Fe/H] and [C/Fe] for Program Stars}

Table 7 reports our determinations of [Fe/H] and [C/Fe] for the BPSII ``strong
G-band'' stars, using the methods described above. Column (1) lists the star name.
In column (2) we list the telescopes involved with the acquisition of the
medium-resolution spectra, using the same coding as in Table 3. The $KP$ and
$GP$ line indices are listed in columns (3) and (4), respectively. Columns (5)
and (6) list estimates of [Fe/H], obtained using the ANNs and regression
procedures described above, respectively.  Columns (7) and (8) list estimates of [C/Fe],
obtained with the ANNs and regression procedures described above, respectively.  
The remaining columns are described below.

\subsection{Spectral Synthesis Calculations for Program Stars}

We have used the procedures described above to obtain spectral synthesis
estimates of [C/Fe] for all but one of our program stars.  In order to obtain these
estimates, we require input model atmosphere parameters, estimated in
the following manner.  Temperatures, which we wish to be on the same scale
as the calibration stars, are obtained from a simple linear regression of
T$_{\rm eff}$ with $(J-K)_0$, using the adopted temperatures of the
atmospheric models of the calibration stars:

\medskip

$$ {\rm T}_{\rm eff} = 6861\; (56) - 3504\; (102)\; (J-K)_0$$
$$ \epsilon = 150~K, \;\;\; {\rm R}^2 = 0.91$$

\medskip
\noindent where the one-sigma estimates of the coefficients are indicated in
parentheses.  

Surface gravities are obtained in a similar manner, relying on the adopted
gravities of the calibration objects:

\medskip

$$\log g = 5.232 \;(0.190) - 6.091\; (0.347)\; (J-K)_0$$
$$\epsilon = 0.51~dex, \;\;\; {\rm R}^2 = 0.71$$

\medskip

Metallicity estimates are taken to be equal to the ANN estimate,
[Fe/H]$_{\rm A}$.  Microturbulance estimates are all set to $\xi = 2.0$ km/s.
The full set of adopted parameters for each program star are listed in column
(9) of Table 7. 

The resulting spectral synthesis estimates of [C/Fe] are listed in column (10)
of Table 7. Above we noted that some stars exhibit such strong CH features
(generally among the cooler stars) that they compromise the sidebands used for
estimation of the $GP$ index; we have indicated these potentially problematic
stars with double colons after the estimate of [C/Fe] in columns (7) and (8) of
Table 7. Also, as above, stars where either the poor quality of the spectrum or
the lack of fit from the synthesis approach led to doubts in the resulting
estimate of [C/Fe]$_{\rm S}$ are indicated with single or double colons next to
the listed estimate of [C/Fe]$_{\rm S}$. 

Column (11) of Table 7 lists our final adopted [C/Fe] determinations for the
program stars, [C/Fe]$_{\rm F}$. In most cases, this is taken to be a straight
average of the reported [C/Fe]$_{\rm A}$ and [C/Fe]$_{\rm S}$ values.  For cases
where we have concerns about the line-index estimates of [C/Fe], we adopt the
spectral synthesis estimates of [C/Fe].

\section{The Distribution of Carbon Abundances for Low-Metallicity Stars}

In Figures 12(a) and 12(b) we show the adopted estimates of [C/Fe] and [C/H] vs.
[Fe/H], respectively, for the ``strong G-band'' stars of BPSII.  This Figure is
quite similar to that obtained by Rossi et al. (1999), but now is based on much
firmer estimates of [Fe/H], [C/Fe], and [C/H].   As is clear from
inspection of this Figure, the stars with [Fe/H] $> -1.0$ are not in fact
carbon-enhanced, although they were flagged as possible cases in BPSII.  The
selection of the possibly carbon-enhanced stars from BPSII was based on a
comparison of the ``strong G-band'' candidates with the median G-band strengths of the
stars contained in a course grid of [Fe/H] and $B-V$.  It appears that this
selection was flawed in the high-metallicity regions.  

The envelope of [C/Fe] clearly becomes more extreme for the stars with the lowest
[Fe/H]. However, as is seen in Figure 12 (b), the upper envelope of [C/H], even
at low [Fe/H], reaches a maximum value of [C/H] $\approx -0.5$. As mentioned in
the introduction, a number of different astrophysical phenomena are likely to be
responsible for the enhancement of carbon at low metallicity. It is interesting
that, independent of [Fe/H], the maximum level of [C/H] that is obtained is
remarkably constant.  If all of the CEMP stars obtained their carbon enhancement
as the result of the transfer of AGB-processed material from a now-deceased
companion, this result places a strong constraint on the level of carbon
enhancement that must be accounted for by models of AGB evolution at low
metallicity.  However, we suspect that it may {\it not} be the case that all of
the CEMP stars can be accounted for by this single process.  It is worth noting
that, for metallicities in the range [Fe/H] $< -2.5$, the distribution of [C/H]
for CEMP stars may be bi-modal, which also suggests that several nucleosynthetic
processes may be involved.

\begin{figure*}[h]
\begin{center}
\includegraphics[angle=270,scale=.5]{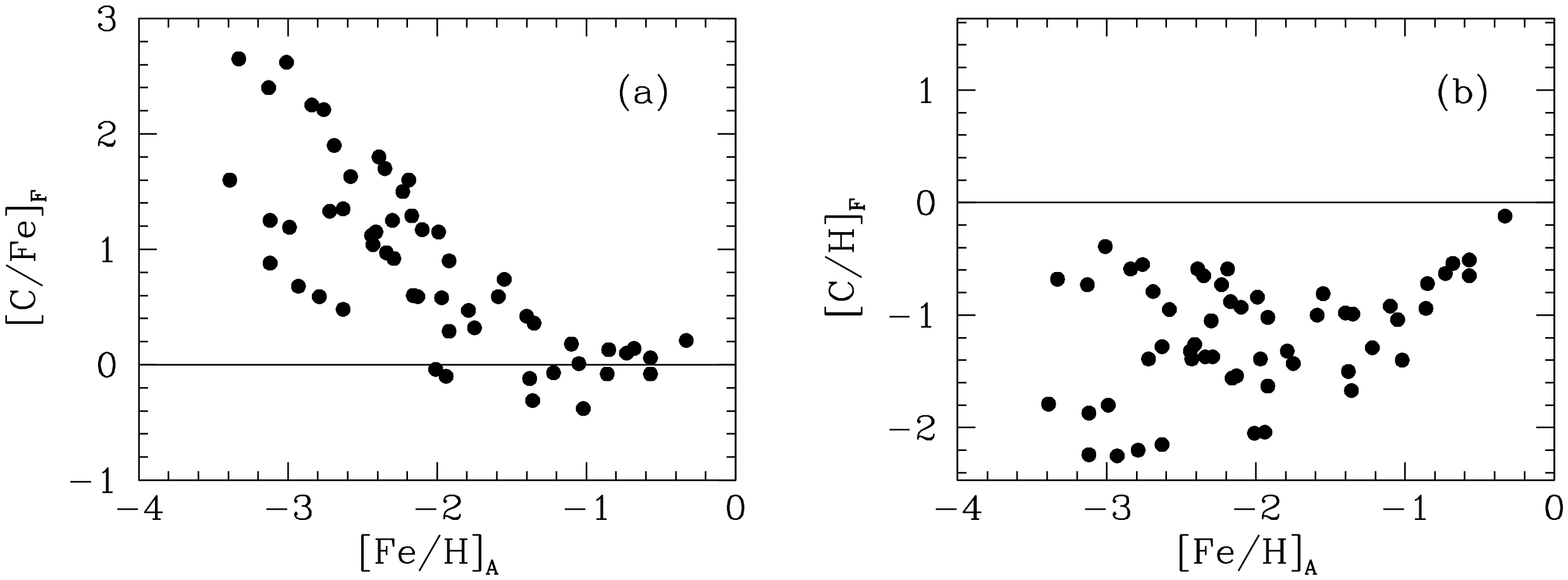}
\end{center}
\caption{Final estimates of the carbon ratios [C/Fe]$_{\rm F}$ and
[C/H]$_{\rm F}$, for the program stars from BPSII, as a function of the ANN
estimate of metallicity [Fe/H]$_{\rm A}$. }
\label{} 
\end{figure*}

\section{Summary and Discussion}

We have described procedures for the estimation of [Fe/H] and [C/Fe] for stars
in the range of abundance $-5.5 \le {\rm [Fe/H]} \le -1.0$, and with
near-infrared colors in the range $0.2 \le (J-K)_0 \le 0.8$. The ANN and
regression approaches exhibit similar errors, although they are slightly worse
than obtained from the spectral synthesis technique. For stars that are either
very cool, or have very large [C/Fe], the synthesis approach is to be preferred,
because of the contamination of the sidebands of the $GP$ index by molecular
carbon lines.  

It is our intention to apply these methods to the determination of [Fe/H] and
[C/Fe] for the large numbers of CEMP stars identified in the HK survey, the HES,
in the sample of C-rich stars identified by Christlieb et al. (2001), and for
carbon-enhanced stars found in SDSS and SEGUE. It may be necessary to
recalibrate some of our estimates of [Fe/H] and [C/Fe] for stars with extremely
strong carbon lines, as well as to better populate the calibration space with
stars of known [Fe/H] and [C/Fe]. This might be accomplished in two ways, either
by including additional stars into the calibrations or by calibrating to a grid
of carbon-enhanced models. Both avenues are presently being pursued. Once this
is accomplished, a more quantitative estimate of the frequency of CEMP stars as
a function of [Fe/H] will be obtained.  This is of clear importance, since the
quoted frequencies of CEMP stars in the literature are based on preliminary
calibrations, and/or partial spectroscopic follow-up. Additional samples for the
investigation of this question are now available from the medium-resolution
stellar spectra obtained during the course of the SDSS; these are presently
being analyzed. The extension of the SDSS, in particular SEGUE, will no doubt
identify many additional CEMP stars, which will be examined in due course.

\bigskip
\bigskip
 
\acknowledgments 
 
The authors would like to thank the directors and support staffs at all of the
observatories that were involved with the acquisition of the spectra used in
this study. We also thank Andy McWilliam and Inese Ivans for sharing their model
stellar atmosphere interpolation code with us.  The authors would also like to
thank an anonymous referee for suggestions that improved our final manuscript,
in particular for pointing out a discrepancy in one of our spectral synthesis plots.
 
S.R. acknowledges partial support from grant
200068/95-4 awarded by The Secretary for Science and Technology, CNPq, Brazil,
as well as from FAPESP. S.R. would also like to thank JINA, as well as the
Chairman and the Department of Physics \& Astronomy at Michigan State University
for their hospitality during her extended visit with them when much of this work was
completed. T.C.B. acknowledges partial support of this work from grants AST 95-29454, AST
00-98508, AST 00-98548, AST 04-06784, and PHY 02-16783, Physics Frontier
Centers/JINA: Joint Institute for Nuclear Astrophysics, awarded by the US
National Science Foundation. C.S. acknowledges partial support from AST 03-07495
awarded by the US National Science Foundation.
 
This work made use of the SIMBAD database, operated at CDS, Strasbourg, France,
as well as NASA's Astrophysics Data System Bibliographic Services.

\clearpage

\begin{deluxetable} {lrcrrccc} 
\tabletypesize{\small} 
\tablewidth{0pt} 
\tablenum{1} 
\tablecaption{Photometric Information for Calibration Stars} 
\tablecolumns{8} 
\tablehead{
\colhead{Star} & \colhead{$V$} & \colhead{$(B-V)$} & \colhead{$J$ } &\colhead{$J-K$} & \colhead{$E(B-V)_A$} & \colhead{$(B-V)_0$} & \colhead{$(J-K)_0$}\\
\colhead{(1)} & \colhead{(2)} & \colhead{(3)} & \colhead{(4)} & \colhead{(5)} & \colhead{(6)} & \colhead{(7)} & \colhead{(8)}
}
\startdata  
BD$+$06:648         &   9.10  &   1.11  &   6.562   (0.020)  &   0.764   (0.027)  &   0.207  &   0.90  &   0.648 \\
BD$+$09:2870        &   9.70  &   0.88  &   7.534   (0.020)  &   0.665   (0.030)  &   0.029  &   0.85  &   0.649 \\
BD$+$09:3223        &   9.25  &   0.56  &   7.760   (0.020)  &   0.483   (0.030)  &   0.076  &   0.48  &   0.440 \\
BD$+$17:3248        &   9.37  &   0.66  &   7.876   (0.056)  &   0.538   (0.059)  &   0.059  &   0.60  &   0.505 \\
BD$+$30:2611        &   9.13  &   1.15  &   6.882   (0.023)  &   0.788   (0.031)  &   0.019  &   1.13  &   0.777 \\
BD$+$34:2476        &  10.06  &   0.38  &   9.077   (0.019)  &   0.266   (0.026)  &   0.014  &   0.37  &   0.258 \\
BD$+$37:1458        &   8.92  &   0.59  &   7.532   (0.020)  &   0.446   (0.027)  &   0.217  &   0.37  &   0.324 \\
BD$-$01:2582        &   9.60  &   0.67  &   8.126   (0.020)  &   0.528   (0.029)  &   0.021  &   0.65  &   0.516 \\
\enddata
\end{deluxetable}

\begin{deluxetable} {lccc} 
\tablewidth{0pt} 
\tablenum{2} 
\tablecaption{Line Index Wavelength Bands} 
\tablecolumns{4} 
\tablehead{\colhead{Line} & \colhead{Line Band} & \colhead{Blue Sideband} & \colhead{Red Sideband}} 
\startdata  
CaII-K6       & 3930.7 -- 3936.7  &  3903.0 -- 3923.0  &  4000.0 -- 4020.0 \\
CaII-K12      & 3927.7 -- 3939.7  &  3903.0 -- 3923.0  &  4000.0 -- 4020.0 \\
CaII-K18      & 3924.7 -- 3942.7  &  3903.0 -- 3923.0  &  4000.0 -- 4020.0 \\
G band ($GP$) & 4297.5 -- 4312.5  &  4247.0 -- 4267.0  &  4363.0 -- 4372.0 \\
\enddata
\end{deluxetable}

\begin{deluxetable} {llrrcccrc} 
\tabletypesize{\small} 
\tablewidth{0pt} 
\tablenum{3} 
\tablecaption{Spectroscopic Observations, and Adopted Abundance and Model Parameters for Calibration Stars} 
\tablecolumns{9} 
\tablehead{
\colhead{Star} & \colhead{Telescope} & \colhead{$~KP$} & \colhead{$~GP$} &\colhead{[Fe/H]} & \colhead{[C/H]} & \colhead{[C/Fe]} & \colhead{Source} & \colhead{Model Parameters\tablenotemark{a}}\\
\colhead{(1)} & \colhead{(2)} & \colhead{~(3)} & \colhead{~(4)} & \colhead{(5)} & \colhead{(6)} & \colhead{(7)} & \colhead{(8)} & \colhead{(9)} 
}
\startdata  
BD$+$06:648       &  K         &   7.28  &   2.39  &  $-$2.33  &  $-$2.23  &  $+$0.10  &        12 &  4500/1.1/-2.3/2.0 \\
BD$+$09:2870      &  k         &   7.75  &   2.91  &  $-$2.29  &  $-$2.90  &  $-$0.61  &        12 &  4600/1.4/-2.3/1.7 \\
BD$+$09:3223      &  k         &   5.51  &   1.09  &  $-$2.19  &  $-$2.70  &  $-$0.51  &        12 &  5350/2.0/-2.2/2.0 \\
BD$+$17:3248      &  ck        &   6.46  &   1.42  &  $-$2.07  &  $-$2.51  &  $-$0.44  &        23 &  5250/1.4/-2.1/1.5 \\
BD$+$30:2611      &  MPk       &  10.93  &   4.61  &  $-$1.24  &  $-$1.90  &  $-$0.66  &        12 &  4170/0.4/-1.2/1.6 \\
BD$+$34:2476      &  k         &   2.88  &   0.61  &  $-$1.98  &  $-$1.94  &  $+$0.04  &        11 &  6320/3.9/-2.0/1.0 \\
BD$+$37:1458      &  k         &   5.74  &   3.28  &  $-$1.96  &  $-$1.73  &  $+$0.24  &     11,12 &  5410/3.6/-2.0/1.0 \\
BD$-$01:2582      &  Lcek      &   5.56  &   6.23  &  $-$2.19  &  $-$1.10  &  $+$0.99  &        12 &  5150/2.5/-2.2/1.2 \\
\enddata

\tablenotetext{a}{The model parameters listed are those adopted based on the literature sources
listed.  For consistency of the spectral synthesis modeling with the predicted [C/Fe] for the
BPSII ``Strong G-Band'' stars in Table 7, we used an input microturbulance value of 2.0 km/sec
for estimation of the [C/Fe]$_{\rm S}$ values reported in this table.}

\tablecomments{Telescope translation codes: A = Anglo Australian 3.8m telescope; C = CTIO 4m telescope;
K = KPNO 4m telescope; I = Isaac Newton 2.5m telescope; L = Las Campanas 2.5m telescope; 
M = McDonald Observatory 2.7m telescope; P = Palomar 5m telescope; c = CTIO 1.5m telescope;
e = ESO 1.5m telescope; k = KPNO 2.1m telescope; s = SSO 2.3m telescope}

\tablerefs{1 = Luck \& Bond (1991); 2 = Ryan et al. (1991);
3 = Smith, Coleman, \& Lambert (1993); 4  = McWilliam et al. (1995);
5 = Shetrone (1996); 6 = Norris et al. (1997a); 7 = Norris et al. (1997b); 8 = Pilachowski et al. (1997);
9 = Bonifacio et al. (1998); 10 = Zacs et al. (1998); 11 = Carretta et al. (2000);
12 = Gratton et al. (2000); 13 = Hill et al. (2000); 14 = Aoki et al. (2001); 15 = Giridhar et al. (2001); 
16 = Preston \& Sneden (2001); 17 = Van Eck et al. (2001); 18 = Aoki et al. (2002a); 19 = Aoki et al. (2002b); 
20 = Aoki et al. (2002c); 21 = Christlieb et al. (2002); 22 = Lucatello et al. (2003); 
23 = Depagne et al. (2002), Cayrel et al. (2004), Spite et al. (2005); 
24 = Honda et al. (2004)} 

\end{deluxetable}

\begin{deluxetable}{lrccccccrc} 
\footnotesize
\tablewidth{0pt} 
\tabletypesize{\scriptsize}
\tablenum{4} 
\tablecaption{Estimates of [Fe/H] and [C/Fe] from ANNs, Regression, and Synthesis for Calibration Stars} 
\tablecolumns{10} 
\tablehead{
\colhead{Star}& 
\colhead{[Fe/H]} & 
\colhead{[Fe/H]$_{\rm A}$} & 
\colhead{[Fe/H]$_{\rm R}$} & 
\colhead{[C/Fe]} & 
\colhead{[C/Fe]$_{\rm A}$} & 
\colhead{[C/Fe]$_{\rm R}$} &  
\colhead{Bins\tablenotemark{a,b,c}} & 
\colhead {[C/Fe]$_{\rm S}$} &  
\colhead{Notes\tablenotemark{d}} \\
\colhead{(1)} & 
\colhead{(2)}    & 
\colhead{(3)}              & 
\colhead{(4)}              & 
\colhead{(5)}    & 
\colhead{(6)}              & 
\colhead{(7)}              &  
\colhead{(8)}                   & 
\colhead{(9)}               &  
\colhead{(10)}                     
}
\startdata   
BD$+$06:648         &   $-$2.33   &   $-$2.39   ($-$0.06)   &   $-$2.34   ($-$0.01)   &   $+$0.10   &   $-$0.55   ($-$0.65)   &   $-$0.48   ($-$0.58)   &   4/3/3   &   $-$0.05   ($-$0.15)    &   X  \\
BD$+$09:2870        &   $-$2.29   &   $-$2.25   ($+$0.04)   &   $-$2.20   ($+$0.09)   &   $-$0.61   &   $-$0.49   ($+$0.12)   &   $-$0.40   ($+$0.21)   &   3/3/4   &   $-$0.75   ($-$0.14)    &      \\
BD$+$09:3223        &   $-$2.19   &   $-$2.21   ($-$0.02)   &   $-$2.06   ($+$0.13)   &   $-$0.51   &   $-$0.66   ($-$0.15)   &   $-$0.75   ($-$0.24)   &   2/3/4   &   $-$0.37:  ($+$0.14)    &      \\
BD$+$17:3248        &   $-$2.07   &   $-$2.18   ($-$0.11)   &   $-$2.02   ($+$0.05)   &   $-$0.44   &   $-$0.70   ($-$0.26)   &   $-$0.76   ($-$0.32)   &   2/3/4   &   $-$0.60   ($-$0.16)    &      \\
BD$+$30:2611        &   $-$1.24   &   $-$1.65   ($-$0.41)   &   $-$1.82   ($-$0.58)   &   $-$0.66   &   $-$0.56   ($+$0.10)   &   $-$0.51   ($+$0.15)   &   4/2/4   &   $-$0.35:: ($+$0.31)    &      \\
BD$+$34:2476        &   $-$1.98   &   $-$1.99   ($-$0.01)   &   $-$2.21   ($-$0.23)   &   $+$0.04   &   $-$0.18   ($-$0.22)   &   $-$0.29   ($-$0.33)   &   1/2/3   & \phs 0.00:: ($-$0.04)    &      \\
BD$+$37:1458        &   $-$1.96   &   $-$1.52   ($+$0.44)   &   $-$1.52   ($+$0.44)   &   $+$0.24   &   $+$0.02   ($-$0.22)   &   $+$0.14   ($-$0.10)   &   2/2/3   &   $+$0.10   ($-$0.14)    &      \\
BD$-$01:2582        &   $-$2.19   &   $-$2.49   ($-$0.30)   &   $-$2.35   ($-$0.16)   &   $+$0.99   &   $+$0.77   ($-$0.22)   &   $+$0.75   ($-$0.24)   &   2/3/2   &   $+$0.80   ($-$0.19)    &      \\
\enddata
\tablenotetext{a}{T$_\mathrm{eff}$ bins:  (1) T$_{\mathrm eff} > 5500 $K; (2) 5000 K $< $ T$_{\mathrm eff} \le 5500$ K; (3) 4500 K $< $ T$_{\mathrm eff} \le 5000$ K; (4) T$_{\mathrm eff} \le 4500$ K}
\tablenotetext{b}{[Fe/H] bins:  (1) [Fe/H] $> -1.0$; (2) $-2.0 < {\rm [Fe/H]} \le -1.0$; (3) $-3.0 < {\rm [Fe/H]} \le -2.0$; (4) [Fe/H] $\le -3.0$}
\tablenotetext{c}{[C/Fe] bins:  (1) [C/Fe] $> +1.0$; (2) $+0.50 < {\rm [C/Fe]} \le +1.0$; (3) $0.0 \le {\rm [C/Fe]} \le +0.5$; (4) [C/Fe] $\le 0.0$}
\tablenotetext{d}{An X in this column indicates that the star was not used in the calibration of the ANNs or the regression estimates for [C/Fe]}
\end{deluxetable}

\begin{deluxetable} {llccccc} 
\tabletypesize{\scriptsize} 
\rotate
\tablewidth{0pt} 
\tablenum{5} 
\tablecaption{Offsets and Scatter in the Estimation of [Fe/H] and [C/Fe] for Calibration Stars} 
\tablecolumns{6} 
\tablehead{
\colhead{Bins} & \colhead{$\Delta \rm{[Fe/H]}_{\rm A}$  ($\sigma \rm{[Fe/H]}_{\rm A}$)} & \colhead{$\Delta \rm{[Fe/H]}_{\rm R}$  ($\sigma \rm{[Fe/H]}_{\rm R}$)} & \colhead{$\Delta \rm{[C/Fe]}_{\rm A}$  ($\sigma \rm{[C/Fe]}_{\rm A}$)} & \colhead{ $\Delta \rm{[C/Fe]}_{\rm R}$  ($\sigma \rm{C/Fe]}_{\rm R}$)} & \colhead{$\Delta \rm{[C/Fe]}_{\rm S}$  ($\sigma \rm{C/Fe]}_{\rm S}$)} \\
\colhead{(1)}  & \colhead{(2)}                                                          & \colhead{(3)}                                                          & \colhead{(4)}                                                          & \colhead{(5)}                                                          & \colhead{(6)}
}
\startdata  

\cutinhead{Global Results}

                                                 &       $-$0.009 (0.230)   &     $+$0.003 (0.264)    &    $-$0.063 (0.259)    &    $-$0.003 (0.289)   &     $+$0.043 (0.245) \\

\cutinhead{T$_{\rm eff}$ Bins}                              

(1) T$_{\rm eff} > 5500$ K                       &       $-$0.104 (0.171)   &     $-$0.152 (0.251)    &    $-$0.265 (0.349)    &    $-$0.336 (0.303)   &     $-$0.254 (0.244) \\
(2) 5000 K $< {\rm T}_{\rm eff} \le 5500$ K      &       $-$0.013 (0.275)   &     $+$0.069 (0.289)    &    $-$0.140 (0.182)    &    $-$0.032 (0.273)   &     $+$0.028 (0.215) \\
(3) 4500 K $\le {\rm T}_{\rm eff} \le 5000$ K    &       $+$0.016 (0.218)   &     $+$0.032 (0.238)    &    $+$0.010 (0.202)    &    $+$0.078 (0.232)   &     $+$0.080 (0.183) \\
(4) T$_{\rm eff} \le 4500$ K                     &       $-$0.016 (0.171)   &     $-$0.132 (0.202)    &    $+$0.051 (0.291)    &    $+$0.094 (0.276)   &     $+$0.209 (0.263) \\

\cutinhead{[Fe/H] Bins}

(1) [Fe/H] $ > -1.0$                             &       $-$0.295 (0.212)   &     $-$0.384 (0.172)    &    $-$0.478 (0.392)    &    $-$0.411 (0.398)   &     $-$0.343 (0.283) \\
(2) -2.0 $< {\rm [Fe/H]} \le -1.0$               &       $+$0.024 (0.245)   &     $+$0.028 (0.289)    &    $-$0.012 (0.196)    &    $+$0.052 (0.198)   &     $+$0.100 (0.235) \\
(3) -3.0 $< {\rm [Fe/H]} \le -2.0$               &       $-$0.018 (0.223)   &     $+$0.050 (0.244)    &    $-$0.082 (0.223)    &    $-$0.042 (0.246)   &     $+$0.036 (0.218) \\
\enddata
\end{deluxetable}

\begin{deluxetable} {lrcrrccc} 
\footnotesize 
\tablewidth{0pt} 
\tablenum{6} 
\tablecaption{Photometric Information for BPSII ``Strong G-Band'' Stars} 
\tablecolumns{8} 
\tablehead{
\colhead{Star} & \colhead{$V$} & \colhead{$(B-V)$} & \colhead{$J$ } &\colhead{$J-K$} & \colhead{$E(B-V)_A$} & \colhead{$(B-V)_0$} & \colhead{$(J-K)_0$}\\
\colhead{(1)} & \colhead{(2)} & \colhead{(3)} & \colhead{(4)} & \colhead{(5)} & \colhead{(6)} & \colhead{(7)} & \colhead{(8)}
}
\startdata  
CS~22172-025    &  \nodata  &  \nodata  &   12.258   (0.028)  &    0.320   (0.037)  &    0.044   &  \nodata  &    0.295   \\
CS~22180-013    &  \nodata  &  \nodata  &   10.588   (0.023)  &    0.411   (0.031)  &    0.025   &  \nodata  &    0.397   \\
CS~22875-026    &  \nodata  &  \nodata  &   13.293   (0.027)  &    0.292   (0.038)  &    0.018   &  \nodata  &    0.282   \\
CS~22877-001    &   12.16   &    0.77   &   10.555   (0.024)  &    0.564   (0.032)  &    0.054   &    0.72   &    0.534   \\
CS~22879-029    &   14.45   &    0.44   &   13.425   (0.026)  &    0.299   (0.041)  &    0.041   &    0.40   &    0.276   \\
CS~22879-144    &   15.13   &    0.40   &   14.108   (0.027)  &    0.247   (0.052)  &    0.038   &    0.36   &    0.226   \\
CS~22880-074    &   13.27   &    0.57   &   11.990   (0.023)  &    0.354   (0.033)  &    0.057   &    0.51   &    0.322   \\
CS~22881-036    &   13.96   &    0.48   &   12.968   (0.026)  &    0.280   (0.040)  &    0.013   &    0.47   &    0.273   \\
\enddata
\end{deluxetable}

\begin{deluxetable}{llcccccccrc} 
\tabletypesize{\scriptsize}
\tablewidth{0pt} 
\tablenum{7} 
\tablecaption{Spectroscopic Observations, Derived Abundances, and Model Parameters for BPSII ``Strong G-Band'' Stars} 
\tablecolumns{11} 
\tablehead{
\colhead{Star} & 
\colhead{Telescope} & 
\colhead{$KP$} & 
\colhead{$GP$} &
\colhead{[Fe/H]$_{\mathrm {A}}$} & 
\colhead{[Fe/H]$_{\mathrm {R}}$} & 
\colhead{[C/Fe]$_{\mathrm {A}}$} & 
\colhead{[C/Fe]$_{\mathrm {R}}$} & 
\colhead{Model Parameters}  & 
\colhead{[C/Fe]$_{\mathrm {S}}$} & 
\colhead{[C/Fe]$_{\mathrm {F}}$} \\
\colhead{(1)}  & 
\colhead{(2)}       & 
\colhead{(3)}  & 
\colhead{(4)}  & 
\colhead{(5)}             & 
\colhead{(6)}              & 
\colhead{(7)}              & 
\colhead{(8)}              & 
\colhead{(9)}               & 
\colhead{(10)}             & 
\colhead{(11)} 
}
\startdata  
CS~22172-025  &   P        &    5.64   &    2.34   &   $-$1.38   &   $-$1.44   &   $-$0.24   &   $-$0.12   &    5830/3.4/-1.4/2.0   &  \phs 0.00:  &  $-$0.12  \\
CS~22180-013  &   Pe       &    9.35   &    6.08   &   $-$0.73   &   $-$0.71   &   $-$0.09   &   $-$0.04   &    5470/2.8/-0.7/2.0   &  $+$0.28     &  $+$0.10  \\
CS~22875-026  &   Le       &    3.61   &    1.82   &   $-$1.92   &   $-$2.07   &   $+$0.17   &   $+$0.32   &    5870/3.5/-1.9/2.0   &  $+$0.42:    &  $+$0.29  \\
CS~22877-001  &   Lcs      &    3.43   &    4.57   &   $-$3.12   &   $-$3.15   &   $+$1.21   &   $+$1.19   &    4990/2.0/-3.1/2.0   &  $+$1.30     &  $+$1.25  \\
CS~22879-029  &   Le       &    2.69   &    3.36   &   $-$2.17   &   $-$2.34   &   $+$1.24   &   $+$1.28   &    5890/3.6/-2.3/2.0   &  $+$1.33     &  $+$1.29  \\
CS~22879-144  &   Le       &    2.80   &    1.79   &   $-$1.79   &   $-$2.10   &   $+$0.53   &   $+$0.68   &    6070/3.9/-1.8/2.0   &  $+$0.42:    &  $+$0.47  \\
CS~22880-074  &   Le       &    3.43   &    5.70   &   $-$2.23   &   $-$2.29   &   $+$1.47:: &   $+$1.39:: &    5700/3.3/-2.2/2.0   &  $+$1.50     &  $+$1.50  \\
CS~22881-036  &   Lce      &    2.55   &    5.33   &   $-$2.19   &   $-$2.36   &   $+$1.85:: &   $+$1.76:: &    5900/3.6/-2.2/2.0   &  $+$1.60     &  $+$1.60  \\
\enddata                                                                                                                                              
\end{deluxetable}


\begin{thebibliography}{} 


\bibitem[]{1036} Allende Prieto, C., Lambert, D. L., \& Asplund, M. 2001, \apj, 556, L63

\bibitem[]{1038} Allende Prieto, C., Lambert, D. L., \& Asplund, M. 2002, \apj, 573, L137

\bibitem[]{1040} Aoki, W., Norris, J.E., Ryan, S.G., Beers, T.C., \& Ando, H. 
2000, \apj, 536, 97

\bibitem[]{1043} Aoki, W., Norris, J.E., Ryan, S.G., Beers, T.C., \& Ando, H. 2002a,
 \apj, 567, 1166 

\bibitem[]{1046} Aoki, W., Ryan, S.~G., 
Tsangarides, S., Norris, J.~E., Beers, T.~C., \& Ando, H. 2003, in Elemental
Abundances in Old Stars and Damped Lyman-{$\alpha$} Systems, 25th meeting of the
IAU, Joint Discussion 15, Sydney, Australia, 15, 19

\bibitem[]{1051} Aoki, W., Norris, J.E., Ryan, S.G., Beers, T.C., \& Ando, H. 2002c, \apjl, 576, 141

\bibitem[]{1053} Aoki, W., Ryan, S.G., Norris, J.E., Beers, T.C., Ando, H., \& Tsangarides, S. 2002b, \apj, 580, 1149

\bibitem[]{1055} Aoki, W., Ryan, S.G., Norris, J.E., Beers, T.C., Ando, H., Iwamoto, N., 
 Kajino, T., Mathews, G.J., \& Fujimoto, M.Y. 2001, \apj, 561, 346 
 

\bibitem[]{1060} Barbuy, B., Cayrel, R., Spite, M., Beers, T.C., Spite, F., Nordstrom, B., Anderson, J.,
 \& Nissen, P.E. 1997, \aap,  317, 63
	
\bibitem[]{1063} Barbuy, B., Spite, M., Spite, F., Hill, V., Cayrel, R., Plez, B.,
\& Petitjean, P. 2005, \aap, 429, 1031

\bibitem[]{1066} Beers, T.C. 1999, in Third Stromlo Symposium: The Galactic Halo, eds. B. Gibson,
 T. Axelrod, \& M. Putman  (ASP: San Francisco), 165, p. 206

\bibitem[]{1069} Beers, T.C., \& Christlieb, N. 2005, ARAA, in press

\bibitem[]{1071} Beers, T.C., Preston, G.W., Shectman, S.A. 1992, \aj, 103, 1987
(BPSII)

\bibitem[]{1074} Beers, T.C., Rossi, S., Norris, J.E., Ryan, S.G., \& Shefler, T. 1999, \aj, 117, 981

\bibitem[]{1076} Bonifacio, P., Monai, S., \& Beers, T.C. 2000, \aj, 120, 2065

\bibitem[]{1078} Bonifacio, P., Beers, T.C., Molaro, P., \& Vladilo, G. 1998, \aap, 332, 672

\bibitem[]{1080} Burstein, D., \& Heiles, C. 1982, \aj, 87, 1165

\bibitem[]{1082} Carretta, E., Gratton, R.G., \& Sneden, C. 2000, \aap, 356, 238	

\bibitem[Castelli \& Kurucz(2003)]{2003IAUS..210P.A20C} Castelli, F., \& 
Kurucz, R.~L.\ 2003, IAU Symposium, 210, 20P 

\bibitem[]{1087} Cayrel, R., Depagne, E., Spite, M., Hill, V., Spite, F., Francois, P., Plez, B., Beers, T.C., Primas, F., Andersen, J., Barbuy, B., Bonifacio, P., Molaro, P., \&  Nordstrom, B. 2004, \aap, 416, 1117

\bibitem[]{1089} Christlieb, N. 2003, Reviews in Modern Astronomy, 16, (Wiley: New York), p.191

\bibitem[]{1091} Christlieb, N., Bessell, M., Beers, T.C., Gustafsson, B., Korn, A., Barklem, P., Karlsson, T.,
 Mizuno-Wiedner, M., \& Rossi, S. 2002, Nature, 419, 904

\bibitem[]{1094} Christlieb, N., Gustafsson, B., Korn, A.J., Barklem, P.S., Beers, T.C., 
 Bessell, M.S., Karlsson, T., \& Mizuno-Wiedner, M. 2004, \apj, 603, 708

\bibitem[]{1123} Cohen, J.G., Shectman, S., Thompson, I., McWilliam, A., Christlieb,
N., Melendez, J., Zickgraf, F.-J., Ramirez, S., \& Swenson, A. 2005, astro-ph/0506745 

\bibitem[]{1097} Cutri, R. M., et al. 2003, 2MASS All Sky Catalog of Point Sources (VizieR Online Data Catalog), 2246

\bibitem[]{1099} Depagne, E., Hill, V., Spite, M., Spite, F., Plez, B., Beers, T.C., Barbuy, B., Cayrel, R.,
 Andersen, J., Bonifacio, P., Francois, P., Nordstrom, B, \& Primas, F. 2002, \aap, 390, 187 

\bibitem[]{1102} Downes, R.A., Margon, B., Anderson, S.F., Harris, H.C., Knapp, G.R., Schroeder, J.,
 Schneider, D.P., York, D.G., Pier, J.R., \& Brinkmann, J. 2004, \aj, 127, 2838	

\bibitem[]{1105} Ellison, S.L., Songaila, A., Schaye, J., \& Pettini, M. 2000, \aj, 120, 1175 

\bibitem[Fitzpatrick \& Sneden(1987)]{1987BAAS...19.1129F} Fitzpatrick, 
M.~J., \& Sneden, C.\ 1987, \baas, 19, 1129 

\bibitem[]{1110} Frebel, A., Aoki, W.,Christlieb, N., Ando, H.,Asplund, M., Barklem, P., Beers, T.C., 
 Eriksson, K., Fechner, C., Fujimoto, M., Honda, S., Kajinao, T., Minezaki, T., Nomoto, K., 
 Norris, J.E., Ryan, S.G., Takada-Hidai, M., Tsangarides, S., \& Yoshii, Y. 2005,
 Nature, 434, 871

\bibitem[]{1115} Giridhar, S., Lambert, D.L., Gonzalez, G., Pandey, G. 2001, \pasp, 113, 519	

\bibitem[]{1117} Gratton, R.G., Sneden, C., Carretta, E., \& Bragaglia, A. 2000, \aap, 354, 169    

\bibitem[]{1119} Hill, V., Barbuy, B., Spite, M., Spite, F., Cayrel, R., Plez, B., Beers, T.C., Nordstrom, B, \& Nissen, P.E. 2000, \aap, 353, 557

\bibitem[]{1121} Holweger, H. 2001, in AIP Conf. Proc. 598, Solar and Galactic Composition:
A Joint SOHO/ACE Workshop, ed. R. F. Wimmer-Schweingruber (New York: AIP), 23

\bibitem[]{1124} Honda, S., Aoki, W., Kajino, T., Ando, H., Beers, T.C., Izumiura, H., Sadakane, K., \& Takada-Hidai, M. 2004, \apj, 607, 474

\bibitem[]{1126} Ibata, R., Lewis, G.F., Irwin, M., Totten, E., \& Quinn, T. 2001, \apj, 551, 294

\bibitem[]{1128} Lodders, K. 2003, \apj, 591, 1220

\bibitem[]{1130} Lucatello, S., Gratton, R., Cohen, J.G., Beers, T.C., Christlieb, N., Carretta, E., \& Ramirez, S. 2003, \aj, 125, 875

\bibitem[]{1132} Lucatello, S., Tsangarides, S., Beers, T.C., Carretta, E., Gratton, R.E.,
 \& Ryan, S.G. 2005a, \apj, 625, 825

\bibitem[]{1135} Lucatello, S., Gratton, R.G., Beers, T.C., \& Carretta, E. 2005b,
\apj, 625, 833 

\bibitem[]{1138} Luck, R.E., \& Bond, H.E. 1991, \apjs, 77, 515

\bibitem[]{1140} Margon, B., Anderson, S.F., Harris, H.C., Strauss, M.A., Knapp, G.R., Fan, X., Schneider, D.P.,
 Vanden Berk, D.E., Schlegel, D.J., Deutsch, E.W., Ivezic, Z., Hall, P.B., Williams, B.F.,
 Davidsen, A.F., Brinkmann, J., Csabai, I., Hayes, J.J.E., Hennessy, G., Kinney, E.K.,
 Kleinman, S.J.; Lamb, D.Q., Long, D., Neilsen, E.H., Nichol, R., Nitta, A., Snedden, S.A.,
 \& York, D.G. 2002, \aj, 124, 1651

\bibitem[]{1146} McWilliam, A., Preston, G.W., Sneden, C., \& Searle, L. 1995, \aj, 109, 2757 

\bibitem[]{1148} Norris, J.E., Ryan. S.G., \& Beers, T.C. 1997a, \apj, 488, 350 

\bibitem[]{1150} Norris, J.E., Ryan, S.G., \& Beers, T.C. 1997b, \apjl, 489, 169

\bibitem[]{1152} Norris, J.E., Ryan, S.G., Beers, T.C., Aoki, W., \& Ando, H. 2002, \apjl, 569, 107 

\bibitem[]{1154} Pettini, M., Madau, P., Bolte, M., Prochaska, J.X., Ellison, S.L., \& Fan, X. 2003, \apj, 594, 695

\bibitem[]{1156} Pilachowski, C., Sneden, C., Hinkle, K., \& Joyce, R. 1997, \aj, 114, 819

\bibitem[]{1158} Preston, G.W., \& Sneden, C.  2001, \aj, 122, 1545

\bibitem[]{1160} Rossi, S., Beers, T.C., \&  Sneden, C. 1999, in Third Stromlo Symposium:  
 The Galactic Halo, eds. B. Gibson, T. Axelrod, \& M. Putman (ASP: San Francisco),
 165, p. 268

\bibitem[]{1164} Ryan, S.G., Norris, J.E., \& Bessell, M.S. 1991, \aj, 102, 303	

\bibitem[]{1166} Schlegel, D.J., Finkbeiner, D.P., \& Davis, M. 1998, \apj, 500, 525

\bibitem[]{1168} Shetrone, M.D. 1996, \aj, 112, 2639    

\bibitem[]{1170} Sivarani, T., Bonifacio, P., Molaro, P., Cayrel, R., Spite, M.,
Spite, F., Plez, B., Andersen, J., Barbuy, B., Beers, T.C., Depagne, E., Hill,
V., François, P., Nordstrom, B., \& Primas, F. 2004, \aap, 413, 1073

\bibitem[]{1174} Smith, V.V., Coleman, H., \& Lambert, D.L. 1993, \apj, 417, 287

\bibitem[Sneden(1973)]{1973ApJ...184..839S} Sneden, C.\ 1973, \apj, 184, 
839 

\bibitem[]{1179} Sneden, C., McWilliam, A., Preston, G.W., Cowan, J.J., Burris, D.L., \&  Armosky, B.J. 1996, \apj, 467, 819
 
\bibitem[]{1181} Sneden, C., Cowan, J.J.,  Lawler, J.E., Ivans, I.I., Burles, S., Beers, T.C., 
 Primas, F., Hill, V., Truran, J.W., Fuller, G.M., Pfeiffer, B., \& Kratz, K.-L. 2003,
 \apj, 591, 936

\bibitem[]{1185} Snider, S., Allende Prieto, C., von Hippel, T., Beers, T.C., Sneden, C.,
 Qu, Y., \& Rossi, S. 2001, \apj, 562, 528

\bibitem[]{1188} Spite, M., Cayrel, R., Plez, B., Hill, V., Spite, F., Depagne, E., Francois, P., Bonifacio, P., 
 Barbuy, B., Beers, T.C., Andersen, J., Molaro, P., Nordstrom, B., \& Primas, F. 2005, \aap, 
 430, 655

\bibitem[]{1192} Steinmetz, M. 2003, in GAIA Spectroscopy: Science and Technology, ed. U. Munari, (ASP: San Francisco), 298, p.381

\bibitem[]{1194} Totten, E.J., \& Irwin, M.J. 1998, \mnras, 294, 1

\bibitem[]{1196} Tsangarides, S., Ryan, S.G., \& Beers 2004, T.C., Mem. S. A. It., 72, 775

\bibitem[]{1198} Van Eck, S., Goriely, S., Jorissen, A., \& Plez, B. 2001, Nature, 412, 793

\bibitem[]{1200} York, D.G. et al. 2000, \aj, 120 1579

\bibitem[]{1202} Zacs, L., Nissen, P.E., \& Schuster, W.J. 1998, \aap, 337, 216

\end{thebibliography}
\end{document}